\documentstyle[onecolumn]{mn}

\newif\ifAMStwofonts

\font\bbbig=msbm10

\font\bbsmall=msbm6
\font\bbtiny=msbm4
\def\Bbb#1{\hbox{\bbbig#1}}

\def\Bs#1{\hbox{\bbsmall#1}}
\def\Bt#1{\hbox{\bbtiny#1}}

\def\sst{\scriptscriptstyle}
\def\be{\begin{equation}}
\def\ee{\end{equation}}
\def\bea{\begin{eqnarray}}
\def\eea{\end{eqnarray}}
\def\pltm{{\raise0.25em\hbox{$\sst +$}\hspace*{-0.525em}\lower0.1em\hbox{$\sst \times$}}}
\def\uu{{\bf u}}
\def\g{{\bf g}}
\def\xx{{\bf x}}
\def\yy{{\bf y}}
\def\zz{{\bf z}}
\def\rr{{\bf r}}
\def\dtilde#1{\stackrel{\scriptscriptstyle \approx}{#1\!}}

\def\temp{1.35}%
\let\tempp=\relax
\expandafter\ifx\csname psboxversion\endcsname\relax
\else
    \ifdim\temp cm>\psboxversion cm
    \else
      \let\temp=\psboxversion
      \let\tempp= 
    \fi
\fi
\tempp
\let\psboxversion=\temp
\catcode`\@=11
\def\psfortextures{
\def\PSspeci@l##1##2{%
\special{illustration ##1\space scaled ##2}%
}}%
\def\psfordvitops{
\def\PSspeci@l##1##2{%
\special{dvitops: import ##1\space \the\drawingwd \the\drawinght}%
}}%
\def\psfordvips{
\def\PSspeci@l##1##2{%
\d@my=0.1bp \d@mx=\drawingwd \divide\d@mx by\d@my
\includegraphics{##1\space}}}%
\def\psforoztex{
\def\PSspeci@l##1##2{%
\special{##1 \space
      ##2 1000 div dup scale
      \number-\psllx\space\space \number-\pslly\space\space translate
}}}%
\def\psfordvitps{
\def\dvitpsLiter@ldim##1{\dimen0=##1\relax
\special{dvitps: Literal "\number\dimen0\space"}}%
\def\PSspeci@l##1##2{%
\at(0bp;\drawinght){%
\special{dvitps: Include0 "psfig.psr"}
\dvitpsLiter@ldim{\drawingwd}%
\dvitpsLiter@ldim{\drawinght}%
\dvitpsLiter@ldim{\psllx bp}%
\dvitpsLiter@ldim{\pslly bp}%
\dvitpsLiter@ldim{\psurx bp}%
\dvitpsLiter@ldim{\psury bp}%
\special{dvitps: Literal "startTexFig"}%
\special{dvitps: Include1 "##1"}%
\special{dvitps: Literal "endTexFig"}%
}}}%
\def\psfordvialw{
\def\PSspeci@l##1##2{
\special{language "PostScript",
position = "bottom left",
literal "  \psllx\space \pslly\space translate
  ##2 1000 div dup scale
  -\psllx\space -\pslly\space translate",
include "##1"}
}}%
\def\psforptips{
\def\PSspeci@l##1##2{{
\d@mx=\psurx bp
\advance \d@mx by -\psllx bp
\divide \d@mx by 1000\multiply\d@mx by \xscale
\incm{\d@mx}
\let\tmpx\dimincm
\d@my=\psury bp
\advance \d@my by -\pslly bp
\divide \d@my by 1000\multiply\d@my by \xscale
\incm{\d@my}
\let\tmpy\dimincm
\d@mx=-\psllx bp
\divide \d@mx by 1000\multiply\d@mx by \xscale
\d@my=-\pslly bp
\divide \d@my by 1000\multiply\d@my by \xscale
\at(\d@mx;\d@my){\special{ps:##1 x=\tmpx cm, y=\tmpy cm}}
}}}%
\def\psonlyboxes{
\def\PSspeci@l##1##2{%
\at(0cm;0cm){\boxit{\vbox to\drawinght
  {\vss\hbox to\drawingwd{\at(0cm;0cm){\hbox{({\tt##1})}}\hss}}}}
}}%
\def\psloc@lerr#1{%
\let\savedPSspeci@l=\PSspeci@l%
\def\PSspeci@l##1##2{%
\at(0cm;0cm){\boxit{\vbox to\drawinght
  {\vss\hbox to\drawingwd{\at(0cm;0cm){\hbox{({\tt##1}) #1}}\hss}}}}
\let\PSspeci@l=\savedPSspeci@l
}}%
\newread\pst@mpin
\newdimen\drawinght\newdimen\drawingwd
\newdimen\psxoffset\newdimen\psyoffset
\newbox\drawingBox
\newcount\xscale \newcount\yscale \newdimen\pscm\pscm=1cm
\newdimen\d@mx \newdimen\d@my
\newdimen\pswdincr \newdimen\pshtincr
\let\ps@nnotation=\relax
{\catcode`\|=0 |catcode`|\=12 |catcode`|
|catcode`#=12 |catcode`*=14
|xdef|backslashother{\}*
|xdef|percentother{
|xdef|tildeother{~}*
|xdef|sharpother{#}*
}%
\def\R@moveMeaningHeader#1:->{}%
\def\uncatcode#1{%
\edef#1{\expandafter\R@moveMeaningHeader\meaning#1}}%
\def\execute#1{#1}
\def\psm@keother#1{\catcode`#112\relax}
\def\executeinspecs#1{%
\execute{\begingroup\let\do\psm@keother\dospecials\catcode`\^^M=9#1\endgroup}}%
\def\@mpty{}%
\def\matchexpin#1#2{
  \fi%
  \edef\tmpb{{#2}}%
  \expandafter\makem@tchtmp\tmpb%
  \edef\tmpa{#1}\edef\tmpb{#2}%
  \expandafter\expandafter\expandafter\m@tchtmp\expandafter\tmpa\tmpb\endm@tch%
  \if\match%
}%
\def\matchin#1#2{%
  \fi%
  \makem@tchtmp{#2}%
  \m@tchtmp#1#2\endm@tch%
  \if\match%
}%
\def\makem@tchtmp#1{\def\m@tchtmp##1#1##2\endm@tch{%
  \def\tmpa{##1}\def\tmpb{##2}\let\m@tchtmp=\relax%
  \ifx\tmpb\@mpty\def\match{YN}%
  \else\def\match{YY}\fi%
}}%
\def\incm#1{{\psxoffset=1cm\d@my=#1
 \d@mx=\d@my
  \divide\d@mx by \psxoffset
  \xdef\dimincm{\number\d@mx.}
  \advance\d@my by -\number\d@mx cm
  \multiply\d@my by 100
 \d@mx=\d@my
  \divide\d@mx by \psxoffset
  \edef\dimincm{\dimincm\number\d@mx}
  \advance\d@my by -\number\d@mx cm
  \multiply\d@my by 100
 \d@mx=\d@my
  \divide\d@mx by \psxoffset
  \xdef\dimincm{\dimincm\number\d@mx}
}}%
\newif\ifNotB@undingBox
\newhelp\PShelp{Proceed: you'll have a 5cm square blank box instead of
your graphics.}%
\def\s@tsize#1 #2 #3 #4\@ndsize{
  \def\psllx{#1}\def\pslly{#2}%
  \def\psurx{#3}\def\psury{#4}
  \ifx\psurx\@mpty\NotB@undingBoxtrue
  \else
    \drawinght=#4bp\advance\drawinght by-#2bp
    \drawingwd=#3bp\advance\drawingwd by-#1bp
  \fi
  }%
\def\sc@nBBline#1:#2\@ndBBline{\edef\p@rameter{#1}\edef\v@lue{#2}}%
\def\g@bblefirstblank#1#2:{\ifx#1 \else#1\fi#2}%
{\catcode`\%=12
\xdef\B@undingBox{
\def\ReadPSize#1{
 \readfilename#1\relax
 \let\PSfilename=\lastreadfilename
 \openin\pst@mpin=#1\relax
 \ifeof\pst@mpin \errhelp=\PShelp
   \errmessage{I haven't found your postscript file (\PSfilename)}%
   \psloc@lerr{was not found}%
   \s@tsize 0 0 142 142\@ndsize
   \closein\pst@mpin
 \else
   \if\matchexpin{\GlobalInputList}{, \lastreadfilename}%
   \else\xdef\GlobalInputList{\GlobalInputList, \lastreadfilename}%
     \immediate\write\psbj@inaux{\lastreadfilename,}%
   \fi%
   \loop
     \executeinspecs{\catcode`\ =10\global\read\pst@mpin to\n@xtline}%
     \ifeof\pst@mpin
       \errhelp=\PShelp
       \errmessage{(\PSfilename) is not an Encapsulated PostScript File:
           I could not find any \B@undingBox: line.}%
       \edef\v@lue{0 0 142 142:}%
       \psloc@lerr{is not an EPSFile}%
       \NotB@undingBoxfalse
     \else
       \expandafter\sc@nBBline\n@xtline:\@ndBBline
       \ifx\p@rameter\B@undingBox\NotB@undingBoxfalse
         \edef\t@mp{%
           \expandafter\g@bblefirstblank\v@lue\space\space\space}%
         \expandafter\s@tsize\t@mp\@ndsize
       \else\NotB@undingBoxtrue
       \fi
     \fi
   \ifNotB@undingBox\repeat
   \closein\pst@mpin
 \fi
\message{#1}%
}%
\def\psboxto(#1;#2)#3{\vbox{%
   \ReadPSize{#3}%
   \advance\pswdincr by \drawingwd
   \advance\pshtincr by \drawinght
   \divide\pswdincr by 1000
   \divide\pshtincr by 1000
   \d@mx=#1
   \ifdim\d@mx=0pt\xscale=1000
         \else \xscale=\d@mx \divide \xscale by \pswdincr\fi
   \d@my=#2
   \ifdim\d@my=0pt\yscale=1000
         \else \yscale=\d@my \divide \yscale by \pshtincr\fi
   \ifnum\yscale=1000
         \else\ifnum\xscale=1000\xscale=\yscale
                    \else\ifnum\yscale<\xscale\xscale=\yscale\fi
              \fi
   \fi
   \divide\drawingwd by1000 \multiply\drawingwd by\xscale
   \divide\drawinght by1000 \multiply\drawinght by\xscale
   \divide\psxoffset by1000 \multiply\psxoffset by\xscale
   \divide\psyoffset by1000 \multiply\psyoffset by\xscale
   \global\divide\pscm by 1000
   \global\multiply\pscm by\xscale
   \multiply\pswdincr by\xscale \multiply\pshtincr by\xscale
   \ifdim\d@mx=0pt\d@mx=\pswdincr\fi
   \ifdim\d@my=0pt\d@my=\pshtincr\fi
   \message{scaled \the\xscale}%
 \hbox to\d@mx{\hss\vbox to\d@my{\vss
   \global\setbox\drawingBox=\hbox to 0pt{\kern\psxoffset\vbox to 0pt{%
      \kern-\psyoffset
      \PSspeci@l{\PSfilename}{\the\xscale}%
      \vss}\hss\ps@nnotation}%
   \global\wd\drawingBox=\the\pswdincr
   \global\ht\drawingBox=\the\pshtincr
   \global\drawingwd=\pswdincr
   \global\drawinght=\pshtincr
   \baselineskip=0pt
   \copy\drawingBox
 \vss}\hss}%
  \global\psxoffset=0pt
  \global\psyoffset=0pt
  \global\pswdincr=0pt
  \global\pshtincr=0pt 
  \global\pscm=1cm 
}}%
\def\psboxscaled#1#2{\vbox{%
  \ReadPSize{#2}%
  \xscale=#1
  \message{scaled \the\xscale}%
  \divide\pswdincr by 1000 \multiply\pswdincr by \xscale
  \divide\pshtincr by 1000 \multiply\pshtincr by \xscale
  \divide\psxoffset by1000 \multiply\psxoffset by\xscale
  \divide\psyoffset by1000 \multiply\psyoffset by\xscale
  \divide\drawingwd by1000 \multiply\drawingwd by\xscale
  \divide\drawinght by1000 \multiply\drawinght by\xscale
  \global\divide\pscm by 1000
  \global\multiply\pscm by\xscale
  \global\setbox\drawingBox=\hbox to 0pt{\kern\psxoffset\vbox to 0pt{%
     \kern-\psyoffset
     \PSspeci@l{\PSfilename}{\the\xscale}%
     \vss}\hss\ps@nnotation}%
  \advance\pswdincr by \drawingwd
  \advance\pshtincr by \drawinght
  \global\wd\drawingBox=\the\pswdincr
  \global\ht\drawingBox=\the\pshtincr
  \global\drawingwd=\pswdincr
  \global\drawinght=\pshtincr
  \baselineskip=0pt
  \copy\drawingBox
  \global\psxoffset=0pt
  \global\psyoffset=0pt
  \global\pswdincr=0pt
  \global\pshtincr=0pt 
  \global\pscm=1cm
}}%
\def\psbox#1{\psboxscaled{1000}{#1}}%
\newif\ifn@teof\n@teoftrue
\newif\ifc@ntrolline
\newif\ifmatch
\newread\j@insplitin
\newwrite\j@insplitout
\newwrite\psbj@inaux
\immediate\openout\psbj@inaux=psbjoin.aux
\immediate\write\psbj@inaux{\string\joinfiles}%
\immediate\write\psbj@inaux{\jobname,}%
\def\toother#1{\ifcat\relax#1\else\expandafter%
  \toother@ux\meaning#1\endtoother@ux\fi}%
\def\toother@ux#1 #2#3\endtoother@ux{\def\tmp{#3}%
  \ifx\tmp\@mpty\def\tmp{#2}\let\next=\relax%
  \else\def\next{\toother@ux#2#3\endtoother@ux}\fi%
\next}%
\let\readfilenamehook=\relax
\def\re@d{\expandafter\re@daux}
\def\re@daux{\futurelet\nextchar\stopre@dtest}%
\def\re@dnext{\xdef\lastreadfilename{\lastreadfilename\nextchar}%
  \afterassignment\re@d\let\nextchar}%
\def\stopre@d{\egroup\readfilenamehook}%
\def\stopre@dtest{%
  \ifcat\nextchar\relax\let\nextread\stopre@d
  \else
    \ifcat\nextchar\space\def\nextread{%
      \afterassignment\stopre@d\chardef\nextchar=`}%
    \else\let\nextread=\re@dnext
      \toother\nextchar
      \edef\nextchar{\tmp}%
    \fi
  \fi\nextread}%
\def\readfilename{\bgroup%
  \let\\=\backslashother \let\%=\percentother \let\~=\tildeother
  \let\#=\sharpother \xdef\lastreadfilename{}%
  \re@d}%
\xdef\GlobalInputList{\jobname}%
\def\psnewinput{%
  \def\readfilenamehook{
    \if\matchexpin{\GlobalInputList}{, \lastreadfilename}%
    \else\xdef\GlobalInputList{\GlobalInputList, \lastreadfilename}%
      \immediate\write\psbj@inaux{\lastreadfilename,}%
    \fi%
    \let\readfilenamehook=\relax%
    \ps@ldinput\lastreadfilename\relax%
  }\readfilename%
}%
\expandafter\ifx\csname @@input\endcsname\relax    
  \immediate\let\ps@ldinput=\input\def\input{\psnewinput}%
\else
  \immediate\let\ps@ldinput=\@@input
  \def\@@input{\psnewinput}%
\fi%
\def\nowarnopenout{%
 \def\warnopenout##1##2{%
   \readfilename##2\relax
   \message{\lastreadfilename}%
   \immediate\openout##1=\lastreadfilename\relax}}%
\def\warnopenout#1#2{%
 \readfilename#2\relax
 \def\t@mp{TrashMe,psbjoin.aux,psbjoint.tex,}\uncatcode\t@mp
 \if\matchexpin{\t@mp}{\lastreadfilename,}%
 \else
   \immediate\openin\pst@mpin=\lastreadfilename\relax
   \ifeof\pst@mpin
     \else
     \edef\tmp{{If the content of this file is precious to you, this
is your last chance to abort (ie press x or e) and rename it before
retexing (\jobname). If you're sure there's no file
(\lastreadfilename) in the directory of (\jobname), then go on: I'm
simply worried because you have another (\lastreadfilename) in some
directory I'm looking in for inputs...}}%
     \errhelp=\tmp
     \errmessage{I may be about to replace your file named \lastreadfilename}%
   \fi
   \immediate\closein\pst@mpin
 \fi
 \message{\lastreadfilename}%
 \immediate\openout#1=\lastreadfilename\relax}%
{\catcode`\%=12\catcode`\*=14
\gdef\splitfile#1{*
 \readfilename#1\relax
 \immediate\openin\j@insplitin=\lastreadfilename\relax
 \ifeof\j@insplitin
   \message{! I couldn't find and split \lastreadfilename!}*
 \else
   \immediate\openout\j@insplitout=TrashMe
   \message{< Splitting \lastreadfilename\space into}*
   \loop
     \ifeof\j@insplitin
       \immediate\closein\j@insplitin\n@teoffalse
     \else
       \n@teoftrue
       \executeinspecs{\global\read\j@insplitin to\spl@tinline\expandafter
         \ch@ckbeginnewfile\spl@tinline
       \ifc@ntrolline
       \else
         \toks0=\expandafter{\spl@tinline}*
         \immediate\write\j@insplitout{\the\toks0}*
       \fi
     \fi
   \ifn@teof\repeat
   \immediate\closeout\j@insplitout
 \fi\message{>}*
}*
\gdef\ch@ckbeginnewfile#1
 \def\t@mp{#1}*
 \ifx\@mpty\t@mp
   \def\t@mp{#3}*
   \ifx\@mpty\t@mp
     \global\c@ntrollinefalse
   \else
     \immediate\closeout\j@insplitout
     \warnopenout\j@insplitout{#2}*
     \global\c@ntrollinetrue
   \fi
 \else
   \global\c@ntrollinefalse
 \fi}*
\gdef\joinfiles#1\into#2{*
 \message{< Joining following files into}*
 \warnopenout\j@insplitout{#2}*
 \message{:}*
 {*
 \edef\w@##1{\immediate\write\j@insplitout{##1}}*
\w@{
\w@{
\w@{
\w@{
\w@{
\w@{
\w@{
\w@{
\w@{
\w@{
\w@{\string\input\space psbox.tex}*
\w@{\string\splitfile{\string\jobname}}*
\w@{\string\let\string\autojoin=\string\relax}*
}*
 \expandafter\tre@tfilelist#1, \endtre@t
 \immediate\closeout\j@insplitout
 \message{>}*
}*
\gdef\tre@tfilelist#1, #2\endtre@t{*
 \readfilename#1\relax
 \ifx\@mpty\lastreadfilename
 \else
   \immediate\openin\j@insplitin=\lastreadfilename\relax
   \ifeof\j@insplitin
     \errmessage{I couldn't find file \lastreadfilename}*
   \else
     \message{\lastreadfilename}*
     \immediate\write\j@insplitout{
     \executeinspecs{\global\read\j@insplitin to\oldj@ininline}*
     \loop
       \ifeof\j@insplitin\immediate\closein\j@insplitin\n@teoffalse
       \else\n@teoftrue
         \executeinspecs{\global\read\j@insplitin to\j@ininline}*
         \toks0=\expandafter{\oldj@ininline}*
         \let\oldj@ininline=\j@ininline
         \immediate\write\j@insplitout{\the\toks0}*
       \fi
     \ifn@teof
     \repeat
   \immediate\closein\j@insplitin
   \fi
   \tre@tfilelist#2, \endtre@t
 \fi}*
}%
\def\autojoin{%
 \immediate\write\psbj@inaux{\string\into{psbjoint.tex}}%
 \immediate\closeout\psbj@inaux
 \expandafter\joinfiles\GlobalInputList\into{psbjoint.tex}%
}%
\def\centinsert#1{\midinsert\line{\hss#1\hss}\endinsert}%
\def\psannotate#1#2{\vbox{%
  \def\ps@nnotation{#2\global\let\ps@nnotation=\relax}#1}}%
\def\pscaption#1#2{\vbox{%
   \setbox\drawingBox=#1
   \copy\drawingBox
   \vskip\baselineskip
   \vbox{\hsize=\wd\drawingBox\setbox0=\hbox{#2}%
     \ifdim\wd0>\hsize
       \noindent\unhbox0\tolerance=5000
    \else\centerline{\box0}%
    \fi
}}}%
\def\at(#1;#2)#3{\setbox0=\hbox{#3}\ht0=0pt\dp0=0pt
  \rlap{\kern#1\vbox to0pt{\kern-#2\box0\vss}}}%
\newdimen\gridht \newdimen\gridwd
\def\gridfill(#1;#2){%
  \setbox0=\hbox to 1\pscm
  {\vrule height1\pscm width.4pt\leaders\hrule\hfill}%
  \gridht=#1
  \divide\gridht by \ht0
  \multiply\gridht by \ht0
  \gridwd=#2
  \divide\gridwd by \wd0
  \multiply\gridwd by \wd0
  \advance \gridwd by \wd0
  \vbox to \gridht{\leaders\hbox to\gridwd{\leaders\box0\hfill}\vfill}}%
\def\fillinggrid{\at(0cm;0cm){\vbox{%
  \gridfill(\drawinght;\drawingwd)}}}%
\def\textleftof#1:{%
  \setbox1=#1
  \setbox0=\vbox\bgroup
    \advance\hsize by -\wd1 \advance\hsize by -2em}%
\def\textrightof#1:{%
  \setbox0=#1
  \setbox1=\vbox\bgroup
    \advance\hsize by -\wd0 \advance\hsize by -2em}%
\def\endtext{%
  \egroup
  \hbox to \hsize{\valign{\vfil##\vfil\cr%
\box0\cr%
\noalign{\hss}\box1\cr}}}%
\def\frameit#1#2#3{\hbox{\vrule width#1\vbox{%
  \hrule height#1\vskip#2\hbox{\hskip#2\vbox{#3}\hskip#2}%
        \vskip#2\hrule height#1}\vrule width#1}}%
\def\boxit#1{\frameit{0.4pt}{0pt}{#1}}%
\catcode`\@=12 
\psfordvips   

\ifoldfss
  \newcommand{\rmn}[1] {{\rm #1}}
  \newcommand{\itl}[1] {{\it #1}}
  \newcommand{\bld}[1] {{\bf #1}}
  \ifCUPmtlplainloaded \else
    \NewTextAlphabet{textbfit} {cmbxti10} {}
    \NewTextAlphabet{textbfss} {cmssbx10} {}
    \NewMathAlphabet{mathbfit} {cmbxti10} {} 
    \NewMathAlphabet{mathbfss} {cmssbx10} {} 
  \fi
  \ifAMStwofonts
    \ifCUPmtlplainloaded \else
      \NewSymbolFont{upmath} {eurm10}
      \NewSymbolFont{AMSa} {msam10}
      \NewMathSymbol{\upi}     {0}{upmath}{19}
      \NewMathSymbol{\umu}     {0}{upmath}{16}
      \NewMathSymbol{\upartial}{0}{upmath}{40}
      \NewMathSymbol{\leqslant}{3}{AMSa}{36}
      \NewMathSymbol{\geqslant}{3}{AMSa}{3E}
      \let\oldle=\le     \let\oldleq=\leq
      \let\oldge=\ge     \let\oldgeq=\geq
      \let\leq=\leqslant \let\le=\leqslant
      \let\geq=\geqslant \let\ge=\geqslant
    \fi
  \fi
\fi 

\ifnfssone
  \newmathalphabet{\mathit}
  \addtoversion{normal}{\mathit}{cmr}{m}{it}
  \addtoversion{bold}{\mathit}{cmr}{bx}{it}
  \newcommand{\rmn}[1] {\mathrm{#1}}
  \newcommand{\itl}[1] {\mathit{#1}}
  \newcommand{\bld}[1] {\mathbf{#1}}
  \def\textbfit{\protect\txtbfit}
  \def\textbfss{\protect\txtbfss}
  \long\def\txtbfit#1{{\fontfamily{cmr}\fontseries{bx}\fontshape{it}%
    \selectfont #1}}
  \long\def\txtbfss#1{{\fontfamily{cmss}\fontseries{bx}\fontshape{n}%
    \selectfont #1}}
  \newmathalphabet{\mathbfit} 
  \addtoversion{normal}{\mathbfit}{cmr}{bx}{it}
  \addtoversion{bold}{\mathbfit}{cmr}{bx}{it}
  \newmathalphabet{\mathbfss} 
  \addtoversion{normal}{\mathbfss}{cmss}{bx}{n}
  \addtoversion{bold}{\mathbfss}{cmss}{bx}{n}
  \ifAMStwofonts
    \ifCUPmtlplainloaded \else
      \UseAMStwoboldmath
      \makeatletter
      \new@mathgroup\upmath@group
      \define@mathgroup\mv@normal\upmath@group{eur}{m}{n}
      \define@mathgroup\mv@bold\upmath@group{eur}{b}{n}
      \edef\UPM{\hexnumber\upmath@group}
      \new@mathgroup\amsa@group
      \define@mathgroup\mv@normal\amsa@group{msa}{m}{n}
      \define@mathgroup\mv@bold\amsa@group{msa}{m}{n}
      \edef\AMSa{\hexnumber\amsa@group}
      \makeatother
      \mathchardef\upi="0\UPM19
      \mathchardef\umu="0\UPM16
      \mathchardef\upartial="0\UPM40
      \mathchardef\leqslant="3\AMSa36
      \mathchardef\geqslant="3\AMSa3E
      \let\oldle=\le     \let\oldleq=\leq
      \let\oldge=\ge     \let\oldgeq=\geq
      \let\leq=\leqslant \let\le=\leqslant
      \let\geq=\geqslant \let\ge=\geqslant
    \fi
  \fi
\fi 

\ifnfsstwo
  \newcommand{\rmn}[1] {\mathrm{#1}}
  \newcommand{\itl}[1] {\mathit{#1}}
  \newcommand{\bld}[1] {\mathbf{#1}}
  \def\textbfit{\protect\txtbfit}
  \def\textbfss{\protect\txtbfss}
  \long\def\txtbfit#1{{\fontfamily{cmr}\fontseries{bx}\fontshape{it}%
    \selectfont #1}}
  \long\def\txtbfss#1{{\fontfamily{cmss}\fontseries{bx}\fontshape{n}%
    \selectfont #1}}
  \DeclareMathAlphabet{\mathbfit}{OT1}{cmr}{bx}{it}
  \SetMathAlphabet\mathbfit{bold}{OT1}{cmr}{bx}{it}
  \DeclareMathAlphabet{\mathbfss}{OT1}{cmss}{bx}{n}
  \SetMathAlphabet\mathbfss{bold}{OT1}{cmss}{bx}{n}
  \ifAMStwofonts
    \ifCUPmtlplainloaded \else
      \DeclareSymbolFont{UPM}{U}{eur}{m}{n}
      \SetSymbolFont{UPM}{bold}{U}{eur}{b}{n}
      \DeclareSymbolFont{AMSa}{U}{msa}{m}{n}
      \DeclareMathSymbol{\upi}{0}{UPM}{"19}
      \DeclareMathSymbol{\umu}{0}{UPM}{"16}
      \DeclareMathSymbol{\upartial}{0}{UPM}{"40}
      \DeclareMathSymbol{\leqslant}{3}{AMSa}{"36}
      \DeclareMathSymbol{\geqslant}{3}{AMSa}{"3E}
      \let\oldle=\le     \let\oldleq=\leq
      \let\oldge=\ge     \let\oldgeq=\geq
      \let\leq=\leqslant \let\le=\leqslant
      \let\geq=\geqslant \let\ge=\geqslant
    \fi
  \fi
\fi 

\ifCUPmtlplainloaded \else
  \ifAMStwofonts \else 
    \def\upi{\pi}
    \def\umu{\mu}
    \def\upartial{\partial}
  \fi
\fi

\title{All-sky search algorithms for monochromatic signals in resonant bar
GW detector data}
\author[J. A. Lobo and M. Montero]
       {J. A. Lobo and M. Montero \\
        Departament de F\'{\i}sica Fonamental, Universitat de Barcelona \\
	Diagonal 647, E-08028 Barcelona, Spain}
\date{28 February 1998}

\pagerange{\pageref{firstpage}--\pageref{lastpage}}
\pubyear{1998}

\begin{document}

\maketitle

\label{firstpage}

\begin{abstract}
In this paper we design and develop several filtering strategies for the
analysis of data generated by a resonant bar Gravitational Wave (GW)
antenna, with the goal to assess the presence (or absence) in them of long
duration monochromatic GW signals, as well as their eventual amplitude and
frequency, within the sensitivity band of the detector. Such signals are
most likely generated in the fast rotation of slightly asymmetric spinning
stars. We shall develop the practical procedures, together with the study
of their statistical properties, which will provide us with useful
information on each technique's performance. The selection of candidate
events will then be established according to threshold-crossing
probabilities, based on the Neyman-Pearson criterion. In particular, it
will be shown that our approach, based on phase estimation, presents better
signal-to-noise ratio than the most common one of pure spectral analysis.
\end{abstract}

\begin{keywords}
Gravitational Waves -- resonant detectors -- monochromatic signals --
data filtering
\end{keywords}

\section{Introduction}
It is generally believed that the most intense Gravitational Waves (GW)
arriving in the Earth from remote sources in the Universe correspond to
very short duration ($\sim$1 millisecond) bursts, generated in the explosion
of a supernova \cite{tho}, or in gamma-ray bursters \cite{roland}. Since
their very first origins, cylindrical bar GW antennae have been applied
to the detection of this sort of events \cite{weber}, and the more modern
cryogenic bars have been used for this purpose, too, with considerably
enhanced sensitivities \cite{nau,as93,alle}: the long decay times of the
oscillations of the bar make it well suited for the measurement of
impulsive, short duration signals \cite{hg,abp}.

It so happens however that some cylindrical GW antennae have been in
continuous operation regime for many consecutive months, even years. This
is the case for example with the {\it Explorer\/} detector, owned and
operated by the {\sl ROG\/} group at Roma (Italy) and installed within
{\sl CERN\/} premises in Geneva (Switzerland) \cite{as93}. Long term
operation naturally provides the appropriate background for a search of
{\it monochromatic\/} signals in the detector data, as requisite long
integration times become available.

Monochromatic signals are most probably generated by the rotation of
asymmetric stars, such as a pulsar or a neutron star. The intensity of the
GWs strongly depends on the amount of asymmetry of the source, and this is
in turn dependent on its equation of state \cite{bg}. Reasonably optimistic
upper bounds on typical star parameters give an extremely weak signal
estimation of $h\/$\,$\sim$\,10$^{-27}$ \cite{tho}, which must be seen
against a noisy background. Clearly, long integration times are required
to reveal this kind of signal.

A systematic search for it must face a practical difficulty which derives
from the fact that the signal is received in the antenna Doppler-shifted due
to the daily and yearly motions of the Earth ---in addition to possible
internal motions within the source if it is e.g. in a binary system. Fourier
analysis of long stretches of data results in high frequency resolution
\cite{kay}, thence in signal spread across several spectrum bins if it is
Doppler shifted. This can naturally cause significant reduction in post-filter
signal-to-noise ratio. The problem is easily overcome if the source position
in the sky is known (or assumed) ahead of time by means of suitable
corrections based on ephemeris calculations. Analyses of this type exist in
the literature: traces of a pulsar in the centre of the supernova SN1987A were
sought in four days of data generated by the 30 metre Garching interferometer
in March 1989 \cite{nie}, and Frasca and La Posta studied almost four years
of data generated by the room temperature bar detector {\sl GEOGRAV\/} in
search for periodic signals from the Galactic Centre and the Large Magellanic
Cloud \cite{flp}. More recently, Mauceli \cite{mau} has looked for
monochromatic GW signals coming from the region of {\it Tuc 47\/} and from
the Galactic Centre in three months of data generated by the cryogenic
detector {\sl ALLEGRO\/} at Luoisiana State University.

A different strategy must of course be used for an all-sky search. The
philosophy of the procedure put forward by Frasca and La Posta \cite{flp}
consists in the construction of a large {\it bank of spectra\/}, taken over
shorter stretches of data such that the frequency resolution in each
individual spectrum be sufficiently low that daily Doppler shifted signals
fit in a {\it single\/} spectral bin. Suitable comparison and averaging are
thereafter applied to the spectra in order to draw conclusions about the
intensity and/or bounds on signals. Astone {\it et al.\/} \cite{as97,as98},
have looked at one year (1991) of data taken by the above referenced
{\it Explorer\/} detector to also perform an all-sky search for monochromatic
sources of GWs. Their method is based upon local maxima identification in
a bank of spectra, followed by {\it close up\/} analyses of frequency peaks
looking for evidence of Doppler shift patterns across the duration of the
entire data stretch.

In this paper we design and develop algorithms for the analysis of data
generated by a resonant bar detector of GWs, in search for monochromatic
signals within the system's sensitive frequency band. We are also interested
in an all-sky search, but adopt a different point of view. Rather than
scanning a bank of spectra, we propose to use a {\it matched filter\/}
technique to estimate both frequency {\it and\/} phase of candidate signals,
then set a threshold, using the Neyman-Pearson criterion, to select those
events which have a given probability of crossing it as a consequence of pure
random noise fluctuations. We have tested our methods in simulations with
real {\it Explorer\/} detector data from 1991, and seen that they perform
very satisfactorily. We plan to apply them to the massive processing of long
stretches of data from the same antenna in a future paper, in order to
provide complementary analyses to the procedures and methods already
reported in \cite{as97} and \cite{as98}.

The article is structured as follows: in section 2 we present a few technical
generalities and set the basic notation conventions. Section 3 is devoted to
a detailed study of a situation in which the signal has a frequency exactly
equal to one of those in the discrete Fourier spectrum of the data
\cite{ERE96}; this corresponds to an idealised situation whose consideration
is methodologically useful, as it allows us to determine the signal's phase,
and to investigate the statistical properties of the filter output; it also
characterizes the main guidelines for the more realistic study in subsequent
sections. In section 4 the method is illustrated with an artificially added
signal to real detector data, which includes the estimation of the noise
spectral density in the presence of such signal. In section 5 we address the
real case, in which the signal frequency no longer exactly matches any of
the discrete samples, so that it {\it leaks\/} across neighbouring spectrum
bins \cite{ERE97}, and also assess the statistical properties of the filter
output \cite{phd}. Finally, in section 6 we apply the method again to real
data with an external control signal added, and show that it works
satisfactorily. The paper closes with a summary of conclusions and future
prospects.

\section{Linear data filtering}

We begin with a review of some fundamental concepts of linear data
processing, fixing also the basic notation which we will be using
throughout this article. 

In the general case, let $\uu(n)$ ($n=0,\ldots,N-1$) be the discrete set
of samples which constitute our experimental data. A linear filter consists
in a discrete set of numbers $\g(n;\mu_i)$ depending on several parameters,
$\mu_i$, which acts on the experimental data as follows:

\be
\yy(\mu_i) = \sum_{n=0}^{N-1} \g(n;\mu_i) \uu(n)
\ee
producing what we shall call {\it the filter output\/}.
It is usually assumed that $\uu(n)$ is the sum of two different contributions:
on the one hand the signal, $\xx(n)$, whose presence we want to assess,
represented by a deterministic function, and on the other hand the noise
$\rr(n)$, a stochastic process:

\be
\uu(n)=\xx(n) + \rr(n). 
\ee
For any choice of parameters, it is appropiate to ask for the filter response
both to the signal, $\yy_{\xx}$, and to the noise, $\yy_{\rr}$, the latter
being a stochastic process, too. The ratio of the mean square values of
these quantities is termed in the literature the output {\it signal-to-noise
ratio\/} ({\it SNR\/}),

\be
\rho \equiv \frac{\yy_{\xx}^2}{<\yy_{\rr}^2>}, \label{eq:rho}
\ee
and it is a measurement of the performance of the filter $\g(n;\mu_i)$.
The theory of the matched filter \cite{hels,papo_sa} precisely determines,
up to a global constant, the functional form which this must have, for
given signal and noise, in order to maximize $\rho$.

In our case we shall assume that the noise $\rr(n)$ can be adequately
modelled by a zero-mean Gaussian and stationary stochastic process, whereas
the signal $\xx(n)$ will be the response of the cryogenic resonant detector
{\it Explorer\/} to a pure monochromatic GW \cite{papi,ERE95},

\be
\xx(n)=A_0 \cos(2\pi f_0 nT + \varphi_0) \label{eq:pmw}.
\ee
Here, $A_0$ is the product of the amplitude by a conversion factor which
defines the detector sensitivity at the frequency of the  gravitational
radiation, $f_g$. This differs from $f_0$ by a constant shift \cite{seg},

\be
f_{ini} =f_g-f_0=900.0267781\ \mbox{Hz},
\ee
introduced by the data-acquisition system of the antenna, with the purpose
to sample the antenna's full bandwidth 27.5088 Hz. The matched filter for
such a signal is then functionally equal to the latter \cite{hels},

\be
\g(n)={\cal B}\cos(2\pi f_0 nT + \varphi_0), \label{eq:mf}
\ee
where ${\cal B}$ is an arbitrary constant.

\section{Non-leaking signals embedded in known spectrum
noise}
\subsection{The signal}
As pointed out in the introduction,
we want to develop in this article a general method
whose operativity does not depend on the existence
of prior information about
the source. 
So, in principle, the 
value of the frequency $f_0$ of the signal 
we can detect must be within the interval
\be
0 \leq f_0 < \frac{1}{2 T},
\ee
where $1/T$ is our Nyquist frequency.
Obviously no search strategy can afford the endless analysis of all
the frequencies in that window, so we shall be forced to select a finite
set of frequencies to scan. 
Nevertheless, the very functional form of the filter shows us that
we shall perform
discrete Fourier transforms (DFTs) in its implementation,
and this defines the set of frequencies which will be searched
in actual practice:  
\be
f T=\frac{k}{N} \hspace {1cm} k \in \{0,\dots,N-1\}.
\label{eq:dft}
\ee
Moreover, for practical reasons,
all the DFTs will be numerically computed using the
fast Fourier transform (FFT) algorithm, a very optimized procedure
which naturally computes at once all spectral components,
with the only restriction that the
number of samples be an exact power of 2. 

In this section, we shall assume that the signal 
is {\it well matched\/} by the spectral template.
By this we mean that the frequency of the signal is in fact
one of those in equation (\ref{eq:dft}),
so that all the signal is in one single bin of the FFT,
with no {\it leakage\/} to the neighbouring ones.
More precisely, we shall be assuming that
\be
f_0 T=\frac{k_0}{N},
\label{eq:fws}
\ee
where $k_0$ is one of $1,\dots,\frac{N}{2} -1$, 
though we do not know which.
We shall disregard the study of any $k_0$ bigger than $N/2$ because
it would be redundant since they represent nothing but negative
frequencies. The value $k_0=0$ is also disregarded since, 
among other considerations,
it represents no wave at all, but a constant signal.
 
Summing up, the target of the present
analysis will be to assess the presence of a signal
\be
\xx(n)=A_0 \cos(2 \pi k_0 n/N +\varphi_0) \label{eq:mgwws}
\ee
in the experimental data series $\uu(n)$,
using a matched filter
\be
\g(n;k,\varphi)={\cal B} \cos(2 \pi k n/N +\varphi). \label{eq:mgwwsf}
\ee
depending on the two unknown paramenters, $k$ and $\varphi$,
which we shall eventually estimate. Besides the advantageous
property of the absence of frequency leakage in the filter output
of such signals,
equation (\ref{eq:mgwws}) shows that $\xx(n)$ is a periodic function
over the entire processed period, because $\xx(N)$ is equal to $\xx(0)$.
In fact, this relationship holds for any sample,
\be
\xx(n+N) =\xx(n) \label{eq:xxN},
\ee 
and it will be a crucial aspect for the developments
which we shall introduce below.

\subsection{The filter performace and the role of ${\cal B}$}
Let us compute the two quantities 
$\yy_{\xx}^2$ and $<\yy_{\rr}^2>$,
in order to evaluate the actual goodness of the filter.
$\yy_{\xx}^2$ is different
from zero only if a signal is really present 
{\it and\/} the value of the parameter $k$ 
matches $k_0$, i.e. the filtered signal
does not leak across different frequency bins,
\be
\yy_{\xx}^2(k,\varphi)=\left[\frac{A_0 {\cal B} N}{2}\right]^2 
\cos^2(\varphi -{\varphi}_0) \; \delta_{k k_0}.
\label{eq:yx2ws}
\ee 
For $<\yy_{\rr}^2(k,\varphi)>$ we have the more complex formula,


\begin{equation}
<\yy_{\rr}^2(k,\varphi)> = \frac{{\cal B}^2 N}{2}
\sum_{n=-(N-1)}^{N-1} R(n)\left(1-\frac{|n|}{N}\right) 
\cos (2\pi k n/N) - {\cal B}^2 \frac{\cos (2\pi k/N-2\varphi)}
{\sin (2\pi k/N)}\,\sum_{n=0}^{N-1} R(n) \sin (2\pi k n/N),
\label{eq:yr2ws}
\end{equation}
where we have introduced the autocorrelation function of the noise,

\be
R(n)\equiv \langle \rr(n'+n) \rr(n') \rangle.
\ee
If we assume\footnote{This, in fact, is
a constraint on the number of samples that we want to filter. We must
set $N$ large enough to sufficiently exceed the correlation time.} 
that $R(n) \sim 0$ for $n \sim N$, it is clear that
the second term shall become negligible in front of the first,
and hence,

\be
<\yy_{\rr}^2(k,\varphi)>\approx {\cal B}^2 \frac{N S(k;N)}{2 T},
\label{eq:yr2wsapp}
\ee
with 

\be
S(k;N) \equiv T \! \sum_{n=-(N-1)}^{N-1} R(n) \left(1-\frac{|n|}{N}\right) 
\cos (2\pi k n/N).
\ee

It can be easily shown that the quantity we have just defined is
the mean value of a periodogramme,

\be
S(k;N) 
=\frac{T}{N}\,
\left\langle\,\left|\sum_{n=0}^{N-1} \rr(n) e^{-i 2\pi k n/N}\right|^2
\,\right\rangle,	  \label{eq:periodogram}
\ee
a well-known way for estimating the power spectral density of the noise at
that particular frequency, based on the Wiener-Khinchine theorem \cite{kay}.

Putting together expressions (\ref{eq:yx2ws}) and (\ref{eq:yr2wsapp})
we finally get for the {\it SNR\/},

\be
\rho=\frac{A_0^2 N T}{2 S(k;N)} \cos^2(\varphi -
{\varphi}_0) \; \delta_{k k_0} \equiv \rho_0 \cos^2(\varphi -
{\varphi}_0) \; \delta_{k k_0}, \label{eq:rhows}
\ee
with 

\be
\rho_0\equiv\frac{A_0^2 N T}{2 S(k;N)},
\ee
the maximum value for $\rho$ we may achieve with the present filter.

The {\it SNR\/} is obviously independent of the constant ${\cal B}$, so
we can freely set it at our will in order to provide $\yy$ with some
advantageous property. Our particular choice is,

\be
{\cal B}(k)\equiv\sqrt{\frac{2 T}{N S(k;N)}},
\ee
the factor that makes $<\yy_{\rr}^2(k,\varphi)>$ equal to one. The
statistical properties of the noise and the linearity of the filter
guarantee that $\yy$ is still Gaussian. Then its probability distribution
will be completely settled once we know its mean $<\yy>$, and its variance,
$\sigma_{\yy}^2$ which in our case coincides with $<\yy_{\rr}^2>$,

\be
\sigma_{\yy}^2\equiv <\yy^2>-<\yy>^2=<\yy_{\rr}^2>=1.
\ee

So, on the one hand, we have forced $\sigma_{\yy}$ to take the same value 
regardless of the particular scanned frequency, and on the other hand, the
mean of $\yy(k,\varphi)$,

\be
<\yy(k,\varphi)>=\sqrt{\rho_0} \cos(\varphi -
{\varphi}_0) \; \delta_{k k_0}, \label{eq:myws}
\ee 
shall be zero when either no signal is present in the data
or, if there is a signal\footnote{In fact, we will put together both cases,
since we shall take the criterion of typifying through $\rho_0=0$ 
any frequency that contains no signal, whether or not there is a signal at
some other value of $k$.}, for any value of $k$ other than $k_0$.
In this way, we have designed a bank of filters
whose outputs corresponding to pure noise are statistically equivalent,
and consequently {\it can be directly compared\/}.

\subsection{Data splitting and averages}

So far, we have implicitly assumed that $N$ represents the total amount of
stored information we have access to or, in other words, that we can analyze
the whole data series in a single filter pass. This is, in many senses,
a rather optimistic assumption. First of all, since we want to process
several months of experimental data, it should not be surprising that the
available (or even existing) computing facilities could not afford such
a calculation. Moreover, the output of any experimental device, like
{\it Explorer\/}, will not be uniform in quality along all the data
acquisition time, and the stationarity of the noise is not preserved over
too long periods of time. Then it could be worse to mix {\it bad\/} data
(those, for instance, with a high level of noise) with {\it good\/} data
in a single analysis, than simply to veto the stretch that we find
unacceptable. But the gaps that we may introduce in rejecting samples of
the experimental set are not the unique discontinuities that we shall find
in the time series, because in such a long term operation of a detector it
is not unlikely that the system suffers sporadic stops. Also the properties
of the physical signal could be not so stable to be satisfactorily fit by
our models along extended periods of time.

We shall thus consider that each series of length $N\/$ is just one among
a set of, say, $M\/$ consecutive\footnote{The relaxation of this condition,
allowing for the existence of missing whole blocks, introduces minor changes
in all the following discussion. So the derived expressions can be easily
adapted to this case.} blocks \cite{as97,vir}. The reasons for the choice
of the particular values of $N\/$ and $M\/$ must not necessarily coincide
in general. In particular, it is possible that there exist several of those
sequences of $N\times M$ data, eventually disconnected, which must be then
processed separately.

So, we shall attach a new label $\alpha$ to each quantity
in order to be able to specify which of the M blocks of N data we refer to:
\be
\yy_{\alpha}(k,\varphi)=
\sqrt{\frac{2 T}{N S(k;N)}} \sum_{n=0}^{N-1} \uu_{\alpha}(n)
\cos(2 \pi k n/N + \varphi)  \hspace{1cm} (\alpha =0,\dots,M-1),
\ee
$\alpha$ is actually a shorthand which simplifies
the notation,
\be
\uu_{\alpha}(n)\equiv \uu(n+\alpha N)\label{eq:u_a}.
\ee

It is obvious, however, that computing $\yy_{\alpha}(k,\varphi)$ for each
$\alpha$ will not change the individual values of $\rho$, as shown by
equation (\ref{eq:rhows}). Our final goal should be then to combine them
in a suitable way which allows us to make the final {\it SNR\/} as high as
possible. The definition (\ref{eq:u_a}) is, in this sense, very revealing
because, when combined with (\ref{eq:mgwws}) and (\ref{eq:xxN}), shows the
most important feature of a non-leaking monochromatic wave: the signal
$\xx_{\alpha}$ in fact does {\it not\/} depend on $\alpha$,

\be
\xx_{\alpha}(n)=A_0\cos(2\pi k_0 n/N+\varphi_0)=\xx(n).
\label{eq:not_a} 
\ee
We consequently see that, if the signal is present, each of these blocks
contains an identical replica of the same stretch of sinusoid in them.
This motivates us to define a new random variable $\zz(k,\varphi)$,

\be
\zz(k,\varphi) \equiv \frac{1}{M} \sum_{\alpha=0}^{M-1} 
\yy_{\alpha}(k,\varphi), \label{eq:zdef}
\ee
whose mean value does not differ from that in equation (\ref{eq:myws}),
but whose variance is reduced as a consequence of this averaging operation.
Before we calculate explicitly this quantity and the value of the new
associated SNR, we are going to focus on the problem of choosing the right
value of the phase parameter of the filter, $\varphi$.

The conceptually simplest method is to compute $\zz(k,\varphi)$ for a lot of
different values of that parameter, then select the best, $\bar{\varphi}$,
i.e., that which gives a larger output after the filtering procedure.

Nevertheless, we do not need to go into such computationally long process,
for the optimum value $\bar{\varphi}$ can be analytically determined as
follows. According to its definition, we may write down $\zz(k,\varphi)$ as

\be
\zz(k,\varphi) = \frac{1}{M} \sqrt{\frac{2 T}{N S(k;N)}} \sum_{\alpha=0}^{M-1} 
\left[ \Re \{\tilde{\uu}_{\alpha} (k)\} \cos \varphi
+\Im \{\tilde{\uu}_{\alpha} (k)\}\sin \varphi \right], 
\ee
where

\be
\tilde{\uu}_{\alpha} (k) \equiv 
\sum_{n=0}^{N-1} \uu_{\alpha}(n) e^{-2\pi i k n/N}
\ee
is the DFT of $\uu_{\alpha}(n)$. So, we define the $\bar{\varphi}$ imposing
a local-maximum condition on $\zz(k,\varphi)$:

\be
{\left. \frac{\partial \zz(k,\varphi)}{\partial 
\varphi} \right|}_{\varphi=\bar{\varphi}} = 0
\label{eq:lmc}
\ee

We thus find\footnote{We can ensure that $\bar{\varphi}$ defined in this way
leads to a maximum and not a minimum of function $\zz(k,\varphi)$
because its second derivative,
$\partial^2_{\varphi} \zz(k,\varphi=\bar{\varphi})=-\bar{\zz}(k)$,
is never positive.}

\be
\tan(\bar{\varphi})= \frac{\displaystyle \sum_{\alpha=0}^{M-1} 
\Im \{\tilde{\uu}_{\alpha} 
(k)\}}{\displaystyle \sum_{\alpha=0}^{M-1} \Re \{\tilde{\uu}_{\alpha} (k)\}}.
\ee

Here it is useful to introduce the two random variables, 
$\Bbb{R}(k)$ and $\Bbb{I}(k)$,

\bea
\Bbb{R}(k) &\equiv& \frac{1}{M} \sqrt{\frac{2 T}{N S(k;N)}}
\sum_{\alpha=0}^{M-1} \Re \{\tilde{\uu}_{\alpha} (k)\}, \\
\Bbb{I}(k) &\equiv& \frac{1}{M} \sqrt{\frac{2 T}{N S(k;N)}}
\sum_{\alpha=0}^{M-1} 
\Im \{\tilde{\uu}_{\alpha} (k)\},
\eea
because they can easily be combined to yield $\zz(k,\bar{\varphi})$,

\be
\bar{\zz}(k) \equiv \zz(k,\bar{\varphi})=
\sqrt{[\Bbb{R}(k)]^2 +[\Bbb{I}(k)]^2},
\ee
where we have defined $\bar{\zz}(k)$ as the best output of the filter at
a given frequency, extending the notation used with $\varphi$. 

The statistical properties of the actual filter output $\bar{\zz}(k)$
will strongly differ from those of $\zz(k,\varphi)$, since
the new random variable is the fruit of a {\it non-linear\/} filtering process.
For instance, its mean is no longer equal to zero, even
if there is no signal in the experimental data, as we shall see.
This is the reason why we did not undertake a
very detailed study of $\zz(k,\varphi)$ in the first place. 

\subsection{Probability distribution of $\bar{\zz}(k)$} 

We shall build the probability density $p(\bar{\zz})$ starting from
$p(\Bbb{R})$ and $p(\Bbb{I})$. Those two auxiliary random variables are
Gaussian by construction, and are statistically independent, as they are the
real and imaginary parts of the Fourier transform of a stationary stochastic
process. Then we only have to know the respective means and variances in
order to complete the information that will fully settle their probability
densities. The mean values of $\Bbb{R}$ and $\Bbb{I}$ can be readily found 
from their definitions and equation (\ref{eq:myws}):

\bea
<\Bbb{R}(k_0)>&=&\sqrt{\rho_0} \cos \varphi_0,\\
<\Bbb{I}(k_0)>&=&\sqrt{\rho_0} \sin \varphi_0
\eea
while for the variances it can be found that both quantities 
are approximately equal, and

\be
\sigma_{\Bs{R}}^2 \approx \sigma_{\Bs{I}}^2 \approx 
\frac{1}{M}, 
\ee
because the correlation time is much shorter
than the duration of the individual series, which is in essence equivalent to
state that, in spite of their consecutiveness, 
they are mutually uncorrelated.

The hypotheses we have made lead us to the following expressions for the
probability density of $\Bbb{R}$ and $\Bbb{I}$:

\bea
p(\Bbb{R})&=&\sqrt{\frac{M}{2 \pi}} e^{-\frac{M}{2} (\Bs{R} 
-\sqrt{\rho_0} \cos \varphi_0)^2}\hspace{3mm} \mbox{and} \\
p(\Bbb{I})&=&\sqrt{\frac{M}{2 \pi}} e^{-\frac{M}{2} (\Bs{I} 
-\sqrt{\rho_0} \sin \varphi_0)^2}.
\eea

Hence $p(\bar{\zz})$ is given, after an integral is evaluated, by

\be
p(\bar{\zz})
=M \bar{\zz} e^{-\frac{M}{2} (\bar{\zz}^2+\rho_0)} 
I_0(M \bar{\zz} \sqrt{\rho_0}),
\label{eq:pzws}
\ee
where we have used one of the integral representations of the
modified Bessel function of order zero \cite{abra}, $I_0(y)$, 

\be
I_0(y)\equiv\frac{1}{2\pi}\int_{0}^{2 \pi}  e^{-y\cos x} dx.
\ee

In the absence of signal, i.e., when $\rho_0=0$, equation (\ref{eq:pzws})
reduces to

\be
p(\bar{\zz}) = M \bar{\zz} e^{-\frac{M}{2} \bar{\zz}^2},
\ee
an expression which explicitly displays the statistical equivalence of all
the frequencies which contain no signal, the property we want to achieve
when we set the value of the constant ${\cal B}$. Moreover, in this case,
$\bar{\zz}$ is Rayleigh distributed or, in other words, $\bar{\zz}^2$
follows a $\chi^2$ distribution with two degrees of freedom \cite{papo_ps}.

\subsection{Mean and variance of $p(\bar{\zz})$. A new SNR.}

We are now interested in the mean and variance of $\bar{\zz}$.
These correspond to the first two moments of the probability distribution
$p(\bar{\zz})$. It appears that a closed analytic expression can be found
for the moments of any order, so we consider it here for completeness.

The $m$-th moment is defined by
\be
<\bar{\zz}^m>\equiv\int_{0}^{\infty} 
M \bar{\zz}^{m+1} e^{-\frac{M}{2} (\bar{\zz}^2+\rho_0)} 
I_0(M \bar{\zz} \sqrt{\rho_0}),
\ee
a calculation that becomes straightforward if one uses
the relationship \cite{grad},
\be
L_{y}(-z) =\frac{1}{\Gamma(y+1)}
\int_{0}^{\infty} x^{y} e^{-(x+z)} I_0(2 \sqrt{x z}) dx,
\label{eq:LI0}
\ee
where $L_{y}(z)$ is the Laguerre function of order $y$, assuming
the normalization condition,
\be
L_{y}(0)=1, \label{eq:Lnorm}
\ee
and $\Gamma(y)$ is Euler's {\it gamma\/} function.
The expectation value of the $m$-th power of $\bar{\zz}$ is then,
\be
<\bar{\zz}^m>=\left(\frac{2}{M}\right)^{\frac{m}{2}}
\Gamma\left( \frac{m}{2} +1 \right) 
L_{\frac{m}{2}}\left(-\frac{M}{2}\rho_0\right).
\label{eq:zm}
\ee

The most relevant moments for our purpose are, as has been said, 
the first and the second. The mean, 
\be
<\bar{\zz}>=\sqrt{\frac{\pi}{2 M}} 
L_{\frac{1}{2}}\left(-\frac{M}{2}\rho_0\right)\label{eq:mz}
\ee 
as we announced, will be different from zero even when $\rho_0$
vanishes, due to the property (\ref{eq:Lnorm}) of Laguerre functions,
\be
\left.<\bar{\zz}>\right|_{\rho_0=0}=\sqrt{\frac{\pi}{2 M}}.
\ee
Nevertheless, the asymptotic behaviour of the Laguerre function
\be
L_{y} (-x)\stackrel{x \gg 1}{\longrightarrow} 
\frac{1}{\Gamma(y+1)} x^y; \label{eq:Lass}
\ee
also shows that $<\bar{\zz}>$,
when a signal is present, approaches the maximum mean value 
which according to (\ref{eq:myws})
the random variables $\yy_{\alpha}(k,\varphi)$ can possibly reach,

\be
<\bar{\zz}> \approx \sqrt{\rho_0}.
\ee

It is worth noting that equation (\ref{eq:Lass}) ensures that
the last expression holds not only when
$\rho_0 \gg 1$ but when $\bar{\rho} \gg 1$, where we have introduced

\be
\bar{\rho}=\frac{M}{2} \rho_0,
\label{eq:brho}
\ee
a quantity that plays the role of the new {\it SNR\/}. The same conclusion
can be obtained after the study of the explicit expressions for
${\bar{\zz}}^2_{\xx}$:

\be
{\bar{\zz}}^2_{\xx} =\rho_0
\ee
and $<{\bar{\zz}}^2_{\rr}>$,

\be
<{\bar{\zz}}^2_{\rr}>=\frac{2}{M}.
\ee

It is relevant to point that, in this case, {\it SNR\/} linearly increases
with the total number of filtered samples, $N\/$\,$\times$\,$M\/$:

\be
\bar{\rho}=\frac{A_0^2 N M T}{4 S(k;N)},
\ee
whereas the more classical procedure of averaging the square of the moduli
of the DFTs leads to

\be
\bar{\rho}_0=\frac{\sqrt{M}}{2} \rho_0=\frac{A_0^2 N \sqrt{M} T}{4 S(k;N)},
\label{rhoclas}
\ee
what means that with our strategy for signals whose frequency is one of the
FFT samples, we enhance by a factor $\sqrt{M}$ the value of $\bar{\rho}$. 

\section{Non-leaking signals embedded in unknown spectrum noise}

\subsection{Replacing $S(k;N)$  \label{se:lu}}

It is almost redundant to say that the operative method we have just
developed requires knowledge of the power spectral density of the noise. 
The aim of this section is the effective substitution of $S(k;N)$ in the
definition of the constant ${\cal B}\/$ by a suitable estimate of this
quantity obtained from the same data series.

Let us begin with a rearrangement of expression (\ref{eq:periodogram}),

\be
S(k;N)=\frac{T}{N}\,\left\langle\left[
\left(\Re\{\tilde{\uu}_{\alpha}(k)\}-\Re\{\tilde{\xx}(k)\}\right)^2
+\left(\Im\{\tilde{\uu}_{\alpha}(k)\}-\Im\{\tilde{\xx}(k)\}\right)^2
\right]\right\rangle.
\label{eq:protoS}
\ee

There is one procedure in this formula that is certainly beyond our control:
we cannot perform the statistical average. Our particular choice will be the
substitution of that operation by a sum over the the entire rank of values
of $\alpha$, because $S(k;N)$, in spite of the formal aspect of
(\ref{eq:protoS}), is {\it independent of the block label\/}. The same
applies upon replacement of $\tilde{\xx}(k)$ (obviously also an unknown
quantity) since $\tilde{\xx}(k)=<\tilde{\uu}(k)_\alpha>$,

\be
\tilde{\xx}(k) \longrightarrow 
\frac{1}{M} \sum_{\alpha=0}^{M-1} \tilde{\uu}_{\alpha}(k).
\ee

Summing up, the random variable we shall use in order to estimate $S(k;N)$
is $\Bbb{S}(k;N)$,


\begin{equation}
\Bbb{S}(k;N) \equiv \frac{T}{N} \frac{1}{M-1} 
\sum_{\alpha=0}^{M-1} \left[
\left(\Re\{\tilde{\uu}_{\alpha}(k)\}
-\frac{1}{M} \sum_{\alpha'=0}^{M-1} \Re\{\tilde{\uu}_{\alpha'}(k)\}\right)^2
+ \left(\Im\{\tilde{\uu}_{\alpha}(k)\}
-\frac{1}{M} \sum_{\alpha'=0}^{M-1} \Im\{\tilde{\uu}_{\alpha'}(k)\}\right)^2
\,\,\right]
\label{eq:BbbS}
\end{equation}
where we have divided by $M$$-1$ and not by $M\/$ because this way we get
an unbiased estimator, i.e.,

\be
<\Bbb{S}(k;N)> = S(k;N). 
\ee

Now we can replace the unknown spectral density by $\Bbb{S}(k;N)$in any
preceding expression, thus obtaining a new filter output $\Bbb{Z}(k)$
which, unlike $\bar{\zz}(k)$, we are able to compute directly from the
raw experimental data $\uu_{\alpha}(n)$. With an analogous procedure to
the one already explained we obtain $p(\Bbb{Z})$ and all its related
quantities, including the corresponding {\it SNR\/}. Instead of starting
from scratch, we shall calculate the probability density of $\Bbb{Z}$ in
two steps, using previous results.

Let us introduce the auxiliary random variable $\Bbb{W}(k)$ 
\be
\Bbb{W}(k) \equiv  \frac{\Bbb{S}(k;N)}{S(k;N)},
\label{eq:W}
\ee
which allows us to define $\Bbb{Z}(k)$ in a simple way,
\be
\Bbb{Z}\equiv \frac{\bar{\zz}}{\sqrt{\Bbb{W}}},
\label{eq:Z}
\ee
thanks to the fact that all the terms containing $S(k;N)$ 
in both random variables {\it mutually\/} cancel out.
Since $\Bbb{W}$ and $\bar{\zz}$ are statistically independent, 
and $p(\bar{\zz})$ was given in the last section, we
have thus reduced the problem to obtaining $p(\Bbb{W})$ and
performing a final integration.

The probability density of $\Bbb{W}$ can be found in
most reference books on Probability \cite{papo_ps},
because it is the arithmetic mean of the squares of 
$2M-2$ zero-mean independent Gaussian variables with unit variances.
So $(2M-2)\, \Bbb{W}$ follows a $\chi^2$ 
distribution with precisely $2M-2$ degrees of freedom,
\be
p(\Bbb{W})=\frac{(M-1)^{M-1}}{\Gamma(M-1)} \Bbb{W}^{M-2} e^{-(M-1) \Bs{W}}.
\ee

\subsection{The actual filter output $\Bbb{Z}$ and its distribution}

The ratio in (\ref{eq:Z}) which we have used for defining $\Bbb{Z}$
is a familiar one in elementary statistics, and is known to follow a
Student's $t$-distribution if $\bar{\zz}$ is a zero-mean Gaussian variable.
Nevertheless, the expression for $p(\bar{\zz})$ is far from a normal
density function, as we have shown in (\ref{eq:pzws}), what compels us to
perform an explicit calculation, leading to the result

\be
p(\Bbb{Z})
=\frac{M \Bbb{Z}}{\left[1+\frac{M}{2M-2} \Bbb{Z}^2\right]^M} 
\exp\left(-\frac{M \rho_0}{2+\frac{M}{M-1} \Bbb{Z}^2}\right)
L_{M-1}\left(-\frac{M^2\Bbb{Z}^2 \rho_0}{4M-4+2M\Bbb{Z}^2}\right),
\label{eq:pZ}
\ee
where once again we have made use of the formula (\ref{eq:LI0}). When no
signal is present in the data at one particular frequency, expression
(\ref{eq:pZ}) reduces to

\be
p(\Bbb{Z})=\frac{M \Bbb{Z}}{\left[1+\frac{M}{2M-2} \Bbb{Z}^2\right]^M}.
\ee

As a matter of fact, $\Bbb{Z}^2$ in this case follows a Fisher's
$F\/$-distribution with 2 and $2M-2$ degrees of freedom, because it is the
ratio of two $\chi^2$ random variables with those degrees of freedom,
respectively.

In order to compute the moments of the density function of $\Bbb{Z}$, the
simplest approach is not to use the final expression of $p(\Bbb{Z})$ but an
intermediate formula 

\be
p(\Bbb{Z})=M \Bbb{Z} \frac{(M-1)^{M-1}}{\Gamma(M-1)} e^{-\frac{M}{2}\rho_0}
\int_{0}^{\infty} \Bbb{W}^{M-1} e^{-\Bs{W}[M-1+\frac{M}{2}\Bs{Z}^2]}
I_0(M\Bbb{Z}\sqrt{\rho_0 \Bbb{W}}) d\Bbb{W},\label{eq:pZi}
\ee
that will avoid the problem of the integration of Laguerre functions with
negative arguments. We thus find 

\be
<\Bbb{Z}^m>
=\left[(M-1)^{\frac{m}{2}+1} 
\frac{\Gamma\left(M-1-\frac{m}{2}\right)}{\Gamma(M)}\right]
\left(\frac{2}{M}\right)^{\frac{m}{2}} \Gamma\left(\frac{m}{2}+1\right)  
L_{\frac{m}{2}} \left(-\frac{M}{2}\rho_0\right),
\ee
where we have chosen a layout that emphasizes the resemblance with the
result corresponding to $<\bar{\zz}^m>$. The term inside the square brackets
approaches unity when $M\gg m$, and then we recover the formula (\ref{eq:zm}).
It is especially interesting to note that, in particular, the newly defined
{\it SNR\/} remains unchanged. Let us split the second order moment of
$p(\Bbb{Z})$

\be
<\Bbb{Z}^2>= \frac{M-1}{M-2}\frac{2}{M}\left(1+\frac{M}{2}\rho_0\right),
\ee
in two terms, namely $<\Bbb{Z}^2_{\rr}>$ and $<\Bbb{Z}^2_{\xx}>$,

\be
<\Bbb{Z}^2>=<\Bbb{Z}^2_{\rr}> + <\Bbb{Z}^2_{\xx}>.
\ee

When no signal is present, $\rho_0=0$, the value of $<\Bbb{Z}^2>$ is merely
due to the response of the noise to the filtering procedure,

\be
<\Bbb{Z}^2_{\rr}>= \frac{M-1}{M-2}\frac{2}{M},
\ee
so we will accordingly assign to the signal the rest of the outcome,

\be
<\Bbb{Z}^2_{\xx}>= \frac{M-1}{M-2} \rho_0,
\ee
and therefore,

\be
\bar{\rho}=\frac{<\Bbb{Z}^2_{\xx}>}{<\Bbb{Z}^2_{\rr}>}=\frac{M}{2}\rho_0.
\ee

\subsection{A practical example \label{se:exam}}

This section is devoted to show the result of such procedure when applied on
a small stretch of data taken by the {\it Explorer\/} in August 3rd of 1991,
and successive days. The starting date was randomly selected since the final
purpose of the present analysis is not extracting conclusions on the presence
of GWs but on the practical performance of the method itself. We have thus
externally introduced a sinusoidal signal with the required absence of
frequency leakage in order to check the ability of the method for revealing
it. The signal, corresponding to a GW with an amplitude of $h_0=10^{-23}$,
was placed at about $921.4$ Hz, near the detector's {\it plus\/} resonance
\cite{as93}. For this particular date the level of noise in the detector was
such that the {\it SNR\/} for this signal was $\rho_0$\,$\sim$\,1/3 for a number
of filtered samples of $N\/$\,=\,131072, a little less than forty minutes.
The signal was therefore completely buried in the noise. The specific value
$N\/$\,=\,131072 may seem arbitrary in this context, but it has a physical
reason: it ensures that no Doppler shift can be observed in the individual
blocks of $N\/$ samples \cite{as97}. Once we set $N\/$ we can pin down the
precise frequency bin which contains the control signal: $k_0$\,=\,50918.

\begin{figure*}
\begin{tabular}{lll}
\hspace*{-0.5 cm} & (a) & (b)\\
\hspace*{-0.5 cm} & \mbox{\psboxto(6.4cm;0cm){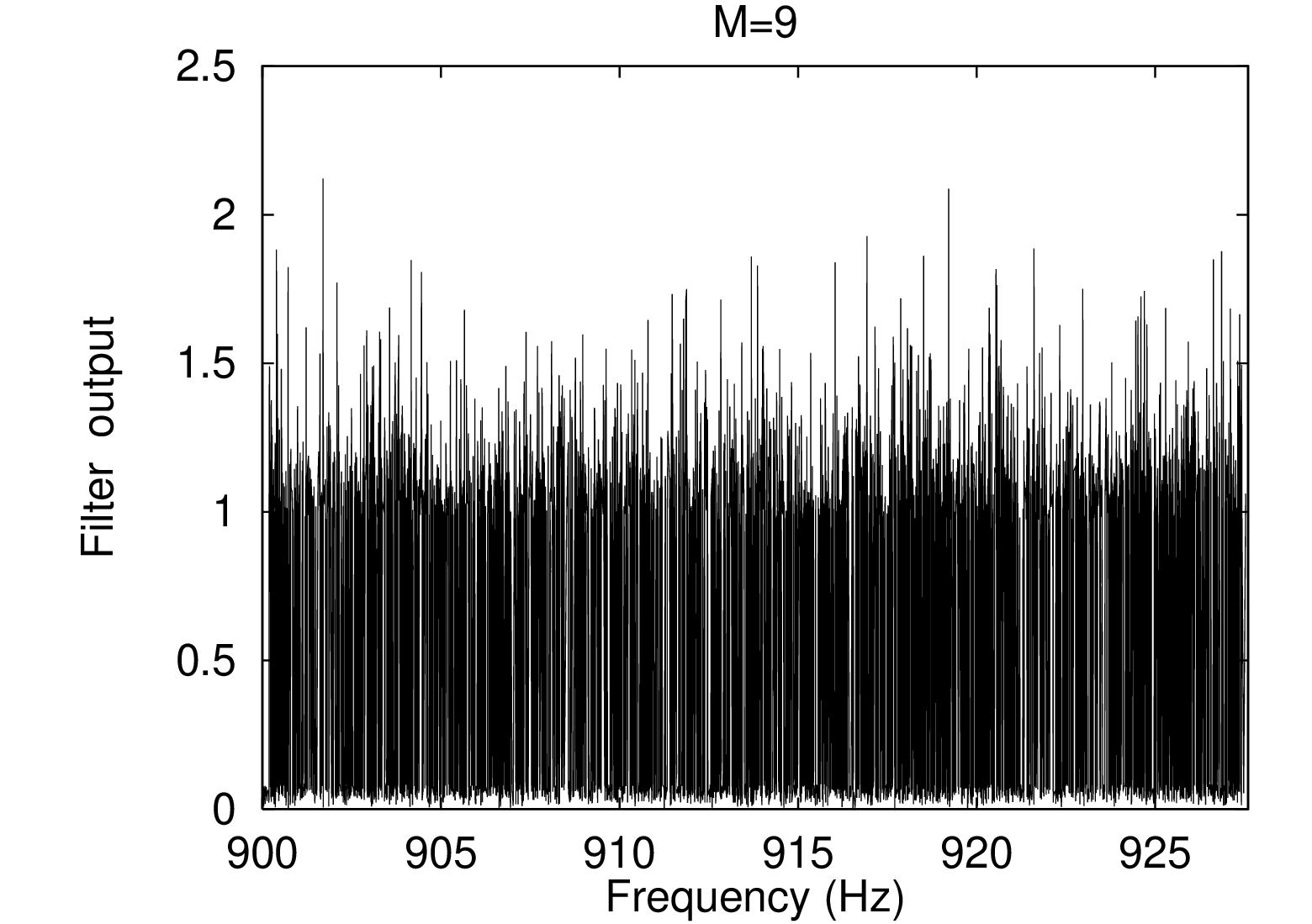}}&
\mbox{\psboxto(6.4cm;0cm){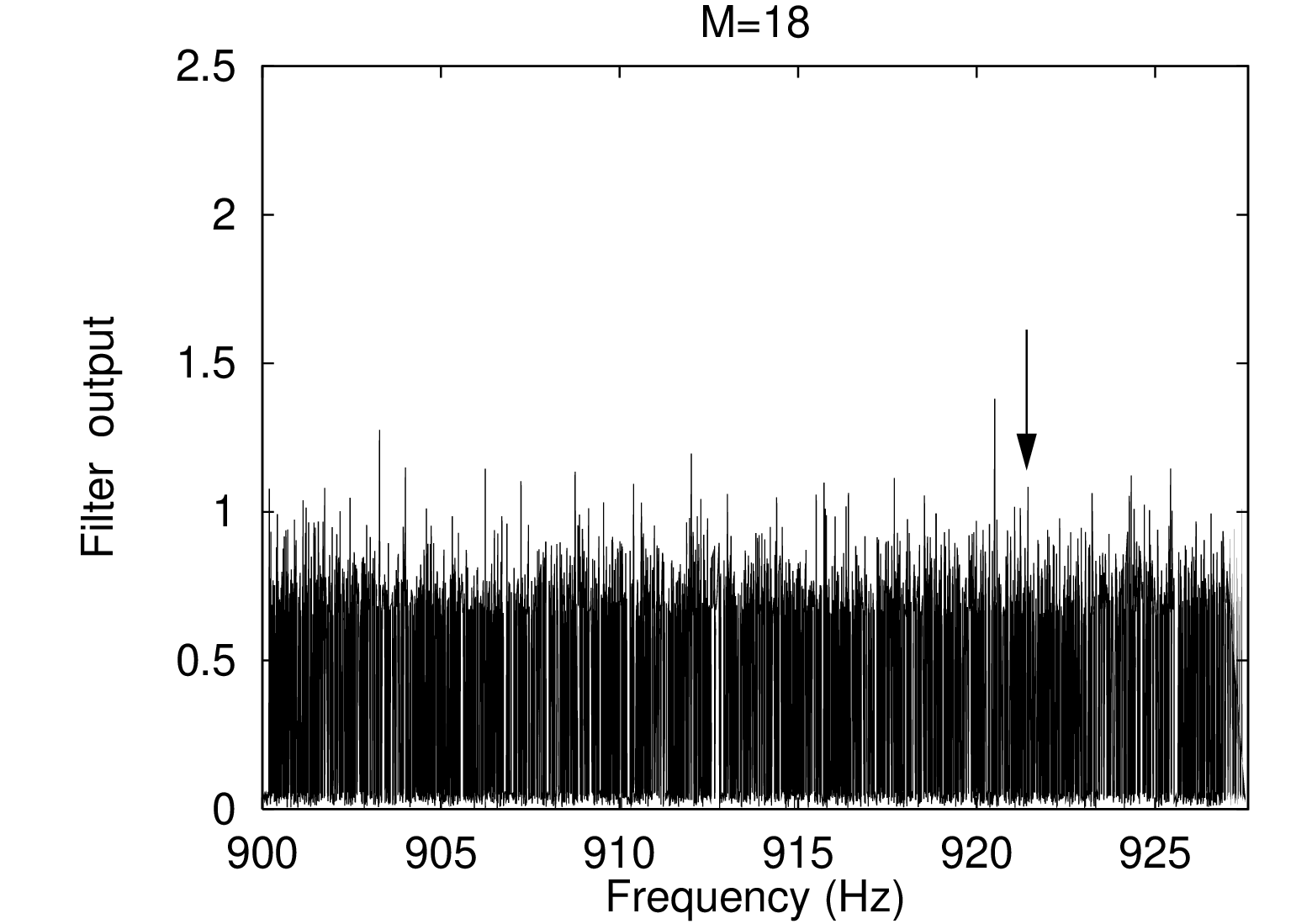}}\\
\hspace*{-0.5 cm} & (c) & (d)\\
\hspace*{-0.5 cm} & \mbox{\psboxto(6.4cm;0cm){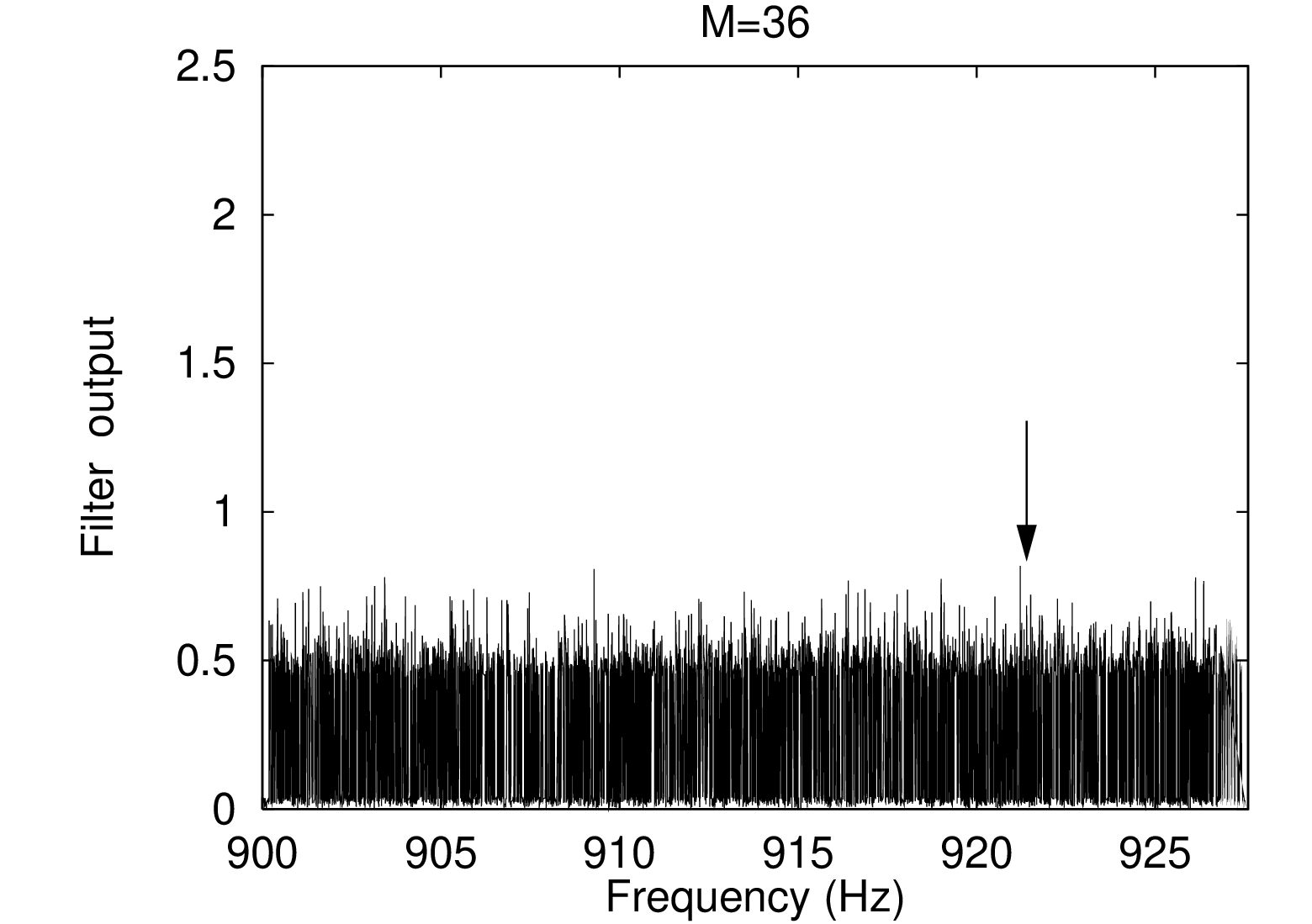}}&
\mbox{\psboxto(6.4cm;0cm){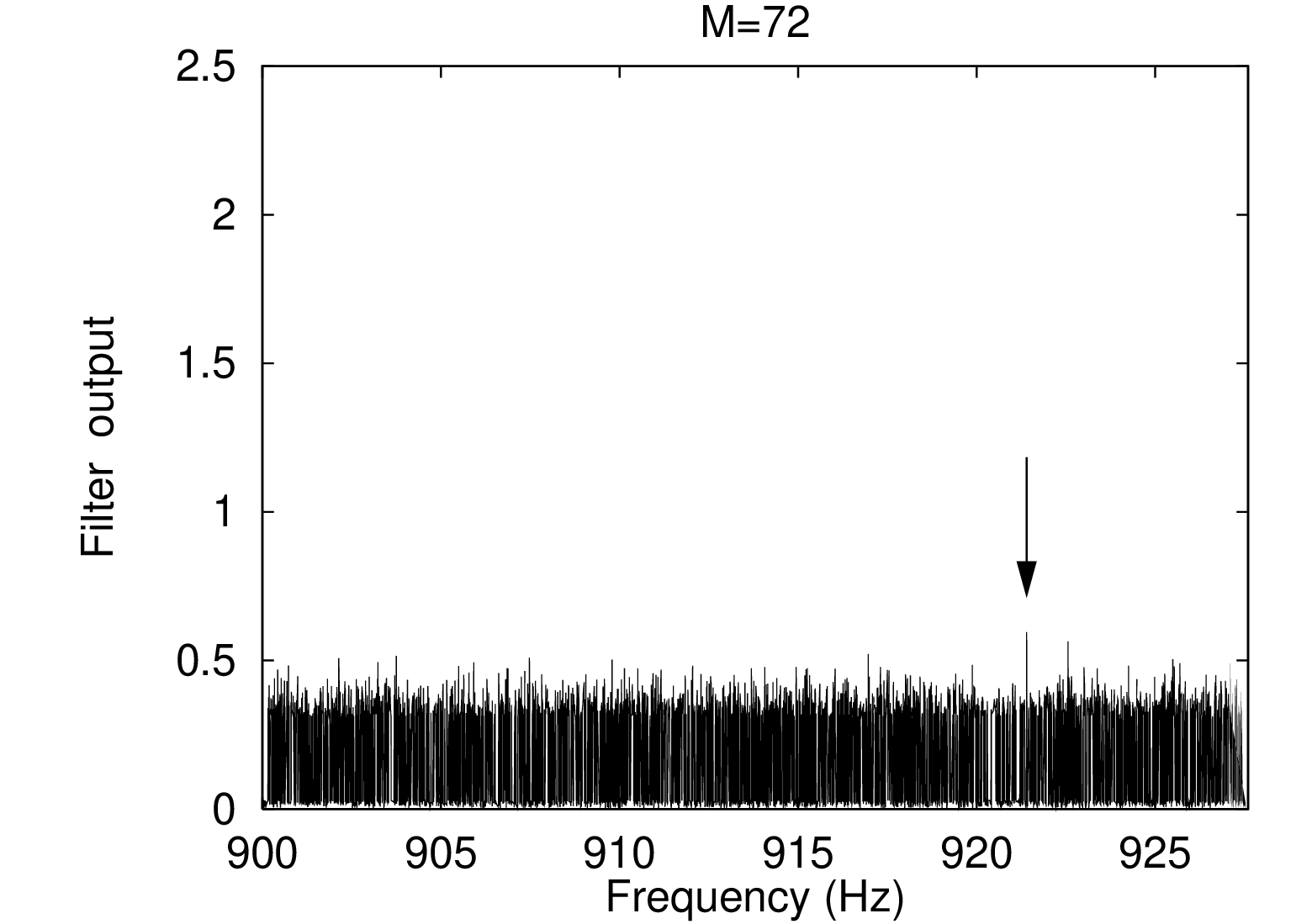}}
\end{tabular}
\caption{Output of the analysis procedure when the filtered data stretches
extend over: (a) six hours, (b) twelve hours, (c) a day and (d) two days.
Every time series begins at the same instant of August 3rd of 1991. The
arrows point to the signal in each of the last three cases. Even though
they are progressively more prominent, only in (d) is the signal the
highest peak. In (c), for example, it is hidden between two taller noise
fluctuations.
\label{fig:zws}}
\end{figure*}
Let us see what happens when we process six hours of data ($M\/$\,=\,9).
Looking at Figure \ref{fig:zws}.a it is by all means impossible to decide
whether the signal is really present or not: {\it SNR\/} has only risen to
a value near unity from the original 1/3 with such few blocks processed. By
increasing $M\/$, i.e., processing longer stretches of data, {\it SNR\/} grows
and the signal becomes progressively more distinct, as we see in Figures
\ref{fig:zws}.b--d, corresponding to half a day, one day and two days of
filtered data, respectively. We must stress at this point that the
theoretical prediction of equation (\ref{eq:brho}) that energy {\it SNR\/}
grows linearly with the number $M\/$ of processed blocks is very accurately
observed in real practice, as we have numerically verified with the plotted
data. The improvement by a factor of $\sqrt{M}\/$ relative to more standard
procedures ---see equation (\ref{rhoclas})--- is thus firmly established not
only in theory but also in actual practice.

\begin{figure*}
\begin{tabular}{lll}
\hspace*{-0.5 cm} & (a) & (b) \\
\hspace*{-0.5 cm} & \mbox{\psannotate{\psboxto(6.4cm;0cm){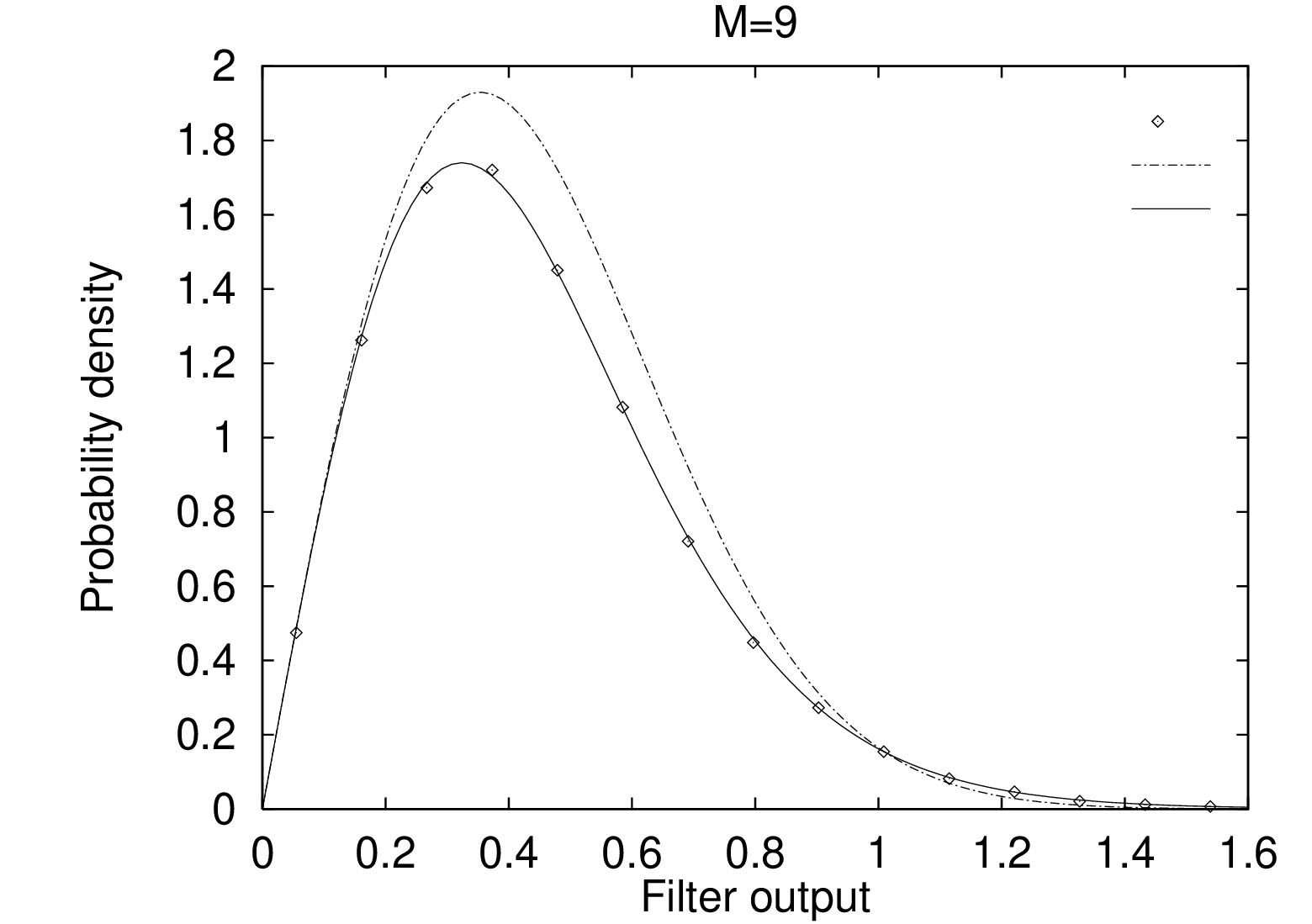}}
{
\at(16\pscm;15.25\pscm){\tiny Experimental}%
\at(19\pscm;14.375\pscm){$\sst p(\bar{\zz})$}%
\at(19\pscm;13.5\pscm){$\sst p(\Bt{Z})$}%
}}&
\mbox{\psannotate{\psboxto(6.4cm;0cm){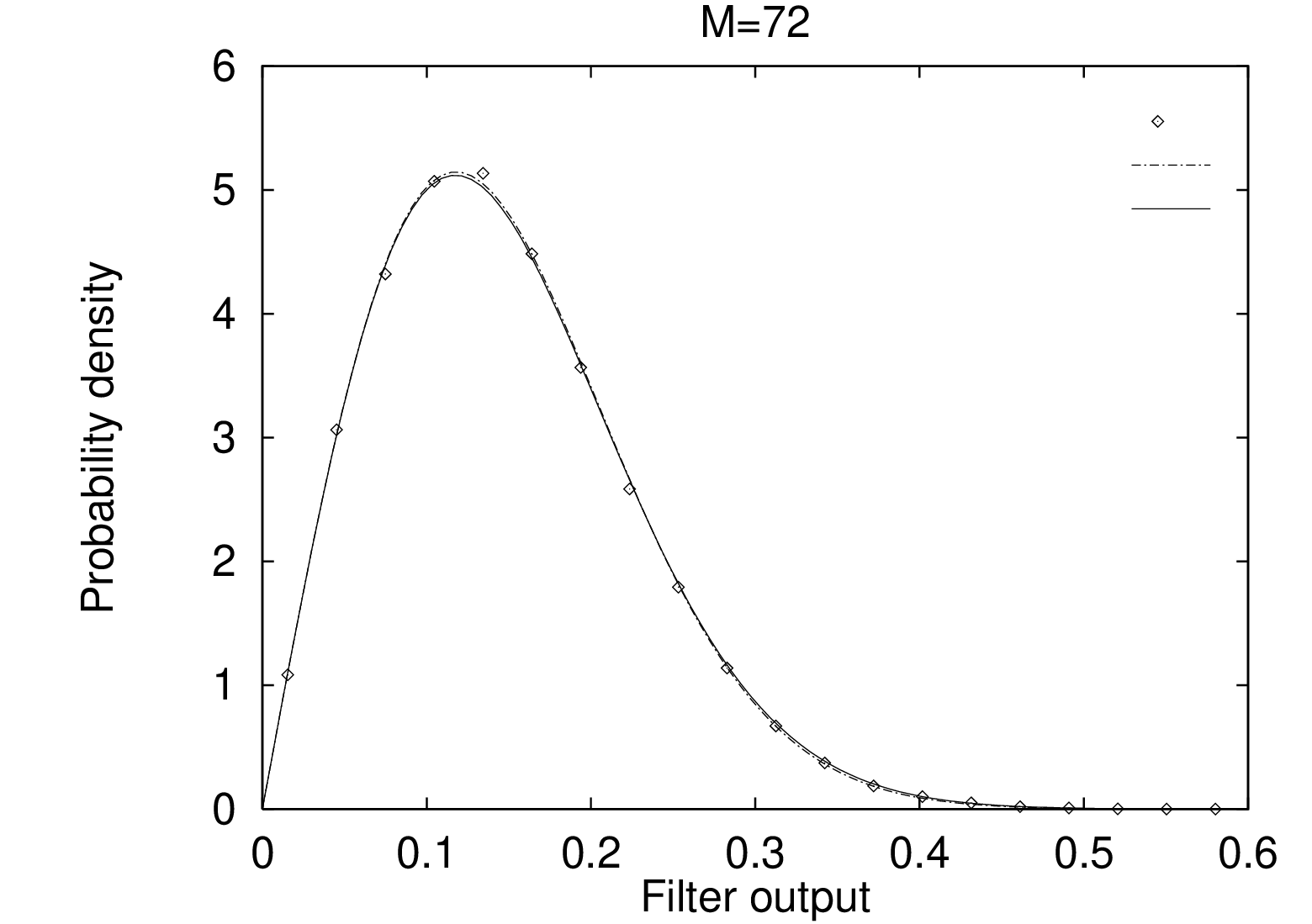}}
{
\at(16\pscm;15.25\pscm){\tiny Experimental}%
\at(19\pscm;14.375\pscm){$\sst p(\bar{\zz})$}%
\at(19\pscm;13.5\pscm){$\sst p(\Bt{Z})$}%
}}
\end{tabular}
\caption{Comparison of the experimental output distribution of $\Bbb{Z}$
with $p(\bar{\zz})$ and $p(\Bbb{Z})$, both in absence of signal, when (a)
$M=9$ and (b) $M=72$.   \label{fig:pzws}}
\end{figure*}

With the output of the filtering process for the values of $k$ other than
$k_0$ we can compute\footnote{
In fact, we do not need to remove $k_0$ (or any other presumed signal)
necessarily before we compute the output distribution, because at most it
is only one point in $2^{16}$.}
the distribution of $\Bbb{Z}$, because when no signal is present it does not
depend on $k\/$. The case $M=9$ is again very interesting because it offers
us the possibility of comparing the experimental distribution with
$p(\bar{\zz})$ and $p(\Bbb{Z})$, thus checking that $\Bbb{Z}$ really follows
the second and not the first ---see Figure \ref{fig:pzws}.a. For higher
values of $M$ the two probability densities converge and become almost
indistinguishable from one another (Figure \ref{fig:pzws}.b). In both
instances, however, the coincidence between theoretical and experimental
distribution is remarkable.

Moreover, the explicit form of $p(\Bbb{Z})$ is very useful, not only in
order to compare it with the experimental one, but to fix a threshold
$\lambda_0$, which shall help us in the task of deciding whether a given
crossing has statistical significance not. We will calculate the {\it error
of the first kind\/}, or false-alarm probability, ${\cal Q}_0$, as a function
of $\lambda_0$,

\be
{\cal Q}_0 \equiv \int_{\lambda_0}^{\infty} p(\Bbb{Z}) d\Bbb{Z}=
\frac{1}{\left[1+\frac{M}{2M-2} \lambda^2\right]^{M-1}}.
\ee
and then we shall set the upper bound depending on the number of {\it false
alarms\/} (i.e. mistakes) we can afford, using the well known Neyman-Pearson
criterion \cite{hels}:

\be
\lambda_0= \sqrt{\frac{2M-2}{M}  
\left(\frac{1}{{\cal Q}_0^{\frac{1}{M-1}}}-1\right)}.
\ee

Table \ref{ta:thres} shows how $\lambda_0$ varies both with respect to
the value of ${\cal Q}_0$ and the number of processed data blocks, $M\/$.
The signal's height, in units of the plots in Figure \ref{fig:zws}, is
0.473, 0.684 and 0.595 for $M\/$\,=\,18, 36 and 72, respectively. It is
therefore above threshold if $M\/$\,=\,72 with false alarm probability
${\cal Q}_0$\,=\,10$^{-5}$, and both if $M\/$\,=\,36 and $M\/$\,=\,72 with
false alarm probability ${\cal Q}_0$\,=\,10$^{-3}$. Even so it cannot be
clearly told from other random fluctuations, as we see in the Figure. This
is because it is very weak, of course. We shall come back to the discussion
of the significance of these thresholds below.

\begin{table}
\caption{Value of the threshold $\lambda_0$ corresponding to different 
choices of ${\cal Q}_0$ for each of the considered block number $M\/$.
\label{ta:thres}}
\begin{minipage}{6.5 cm}
\begin{tabular}{|c|c|c|c|c|}
${\cal Q}_0$ & $M=9$ &$M=18$ &$M=36$ & M=72 \\ \hline
$10^{-3}$    & 1.56  & 0.97  & 0.65  & 0.45 \\
$10^{-5}$    & 2.39  & 1.35  & 0.87  & 0.59 \\ \hline
\end{tabular}
\end{minipage}
\end{table}

\section{Leaking signals embedded in known spectrum noise}
\label{se:leak}

\subsection{A leaking signal}

We are going to start this section considering the effects that the presence
of a general frequency signal in the data may produce in the results we have
exposed in the preceding sections. In particular, we shall study the new
statistical properties of the random variables $\yy_{\alpha}(k,\varphi)$,
that determine the characteristics of $\Bbb{R}$ and $\Bbb{I}$, and
consequently of $\bar{\zz}$. 

So, in the following we shall relax the condition (\ref{eq:fws})

\be
f_0 T =\frac{k_0 + \epsilon_0}{N}
\ee 
by introducing the real parameter $\epsilon_0$

\be
-\frac{1}{2}\leq\epsilon_0 <\frac{1}{2},
\ee
whose effective consequence is that $\xx_{\alpha}(n)$ is no longer
independent of $\alpha$, which appears in the form of an accumulative
phase shift whenever $\epsilon_0$ is different from zero:

\be
\xx_{\alpha}(n)=
A_0 \cos(2 \pi [k_0 +\epsilon_0] n/N + 2 \pi \alpha \epsilon_0 + \varphi_0).
\ee

While it is true that the frequency remainder also lowers the maximum filter
output, due to the spectral leakage of the signal,
\be
<\yy_{\alpha}(k_0,\varphi)>=\sqrt{\rho_0} 
\frac{\sin (\pi \epsilon_0)}{N \sin (\pi \epsilon_0/N)}
\cos \left(\pi \epsilon_0 \left[1- \frac{1}{N}\right]+\varphi_0-\varphi
+2\pi \alpha \epsilon_0\right);
\ee
it is nevertheless the block dependence that damages the filtering procedure,
because it is responsible of the sinc-like behaviour of the mean values of
$\Bbb{R}(k_0)$ and $\Bbb{I}(k_0)$, when considered as functions of $M\/$:

\bea
<\Bbb{R}(k_0)>&=&\sqrt{\rho_0} 
\frac{\sin (\pi \epsilon_0 M)}{N M\sin (\pi \epsilon_0/N)}
\cos \left(\pi \epsilon_0 \left[M- \frac{1}{N}\right]+\varphi_0\right),\\[1 em]
<\Bbb{I}(k_0)>&=&\sqrt{\rho_0} 
\frac{\sin (\pi \epsilon_0 M)}{N M\sin (\pi \epsilon_0/N)}
\sin \left(\pi \epsilon_0 \left[M- \frac{1}{N}\right]+\varphi_0\right).
\eea

This fact turns the search strategy not so robust as desired, in the sense
that for a given $\epsilon_0$ other than zero, there always exists a value
$M_0$ ($\sim 1/\epsilon_0$) for $M\/$, above which the {\it SNR\/} decreases
noticeably. The frequency band where the analysis method works efficiently
thus decreases with the number of averaged blocks, a very undesirable
feature. We now investigate how this problem can be addressed.

\subsection{Phase-varying filter}

As already stated, since the origin of the problem is a carried over phase,
we shall solve it introducing a new block-dependent filter with one more
parameter, $\epsilon$:

\be 
\g_{\alpha}(n;k,\varphi,\epsilon)=
\sqrt{\frac{2 T}{N S(k;N)}}\cos (2 \pi k n /N + \varphi+2 \pi \alpha \epsilon).
\ee
with the purpose to compensate for the phase shift. Starting from this new
$\g_{\alpha}(n;k,\varphi,\epsilon)$, we can calculate the value of each
$\yy_{\alpha}(k,\varphi,\epsilon)$,

\be
\yy_{\alpha}(k,\varphi,\epsilon)=\sqrt{\frac{2 T}{N S(k;N)}} 
|\tilde{\uu}_{\alpha}(k)| 
\cos(\Phi_{\alpha}(k)-\varphi-2\pi \alpha \epsilon),
\label{eq:y_a}
\ee
where the following notation has been used:

\be
\tilde{\uu}_{\alpha}(k)=|\tilde{\uu}_{\alpha}(k)| 
e^{i \Phi_{\alpha}(k)}.
\ee

Through a definition formally identical to (\ref{eq:zdef}), we shall
establish $\zz(k,\varphi,\epsilon)$. Once again it is possible to obtain
$\bar{\varphi}$ using a local-maximum condition, like that in (\ref{eq:lmc}),

\be
{\left.\frac{\partial \zz(k,\varphi,\epsilon)}{\partial \varphi} 
\right|}_{\varphi=\bar{\varphi}}=0,
\; \Longrightarrow \; 
\tan(\bar{\varphi})= \frac{\displaystyle \sum_{\alpha=0}^{M-1} 
|\tilde{\uu}_{\alpha}(k)| 
\sin(\Phi_{\alpha}(k)-2\pi \alpha \epsilon)}{\displaystyle 
\sum_{\alpha=0}^{M-1} 
|\tilde{\uu}_{\alpha}(k)| 
\cos(\Phi_{\alpha}(k)-2\pi \alpha \epsilon)}.
\ee

The value of $\bar{\varphi}$ leads now to the following expression for 
$\zz(k,\bar{\varphi},\epsilon)$,

\be
\zz(k,\bar{\varphi},\epsilon)=\frac{1}{M}\sqrt{\frac{2 T}{N S(k;N)}}
|\dtilde{\uu}(k;\epsilon)|,
\ee
a relationship that involves a new quantity, 
$\dtilde{\uu}(k;\epsilon)$,
which formally is also a DFT,
\be
\dtilde{\uu}(k;\epsilon)\equiv
\sum_{\alpha=0}^{M-1} \tilde{\uu}_{\alpha} (k) e^{-2 \pi i \alpha \epsilon}.
\label{zoom}
\ee

The template for $\epsilon$, just like in the case of the frequency grid, 
will be dictated by the convenience of the use of the FFT algorithm in the
computation of $\dtilde{\uu}(k;\epsilon)$. We shall therefore estimate
$\epsilon_0$ within the following discrete rank of values\footnote{
As usual, any value for q equal or larger than $M'/2$ is related with a
negative $\epsilon$.} of $\epsilon$:

\be
\epsilon=\frac{q}{M'} \hspace{1cm} q \in \{0,\ldots,M'-1\},
\label{zoomspan}
\ee
where, in principle, $M'$, by the way an exact power of 2, must not
necessarily coincide with $M$. $M'$ has to be greater than $M\/$ if we do
not want to waste available information, but on the other hand, it seems
that larger and larger values of $M'$ should produce an endless increment 
of precision in the estimation of $\epsilon_0$. Nevertheless, as we will
show in the next section, $M'$ should be kept as low as possible.

Equation (\ref{zoom}), together with the span condition (\ref{zoomspan}),
bears a considerable formal resemblance with the so called {\it zoom
transform\/} \cite{yip,hung,dewild}, in which the {\it twiddle\/} factor
is now identically equal to one. It must however be stressed that, in the
present context, ours is an {\it interpolation\/} formula rather than
a frequency resolution algorithm. Our formalism can thus be naturally
extended with the purpose to refine the spectral resolution, simply taking
$M'$\,=\,$M\/$, i.e., data sets of $N\times M'\/$ points, where all possible
values of $q\/$ must be considered, for a selected choice of frequency bins,
$k\/$.

\subsection{Statistical properties of $\bar{\zz}$}

The mechanism for finding $\bar{q}$ is then very simple. 
We must compute all the $\zz(k,\bar{\varphi},q/M')$ and select
that $q$ which gives the largest output, just defining 
$\bar{\zz}(k)=\zz(k,\bar{\varphi},\bar{q}/M')$, i.e.,
\be
\bar{\zz}(k)=
\begin{array}[t]{c}
\mbox{max} \\
{\sst q\in \{0,\ldots,M'-1\}}
\end{array}
\left\{\zz(k,\varphi,q/M')\right\}.
\ee 
When no signal is present at some frequency,
it can be proved that again,
\be
p(\zz)=M \zz e^{-\frac{M}{2}\zz^2},
\ee
no matter the choices for $k$ and $q$. So, in that case,
the probability density of $\bar{\zz}(k)$, since it is 
the maximum of $M'$ equally distributed random variables \cite{papo_ps}, 
is given by 
\be
p(\bar{\zz})=M M' \bar{\zz} e^{-\frac{M}{2} \bar{\zz}^2} 
\left[1-e^{-\frac{M}{2} \bar{\zz}^2}\right]^{M'-1}.
\label{eq:pzz}
\ee
We have not been able to compute $<\bar{\zz}>$ 
other than in the form of an alternating finite series, which is almost useless
for obtaining generic conclusions about it. Instead of 
the mean value of $\bar{\zz}$ we shall compute its most probable value. 
The function $p(\bar{\zz})$ 
reaches its maximum (when $M' \gg \sqrt{e}$) for
\be
\bar{\zz} \approx \sqrt{\frac{2}{M} \ln M'},
\ee
a quantity that decreases as $M$ increases, and when $M'$ decreases. 
This shows the convenience of setting $M'$ as the first exact power of 2
greater than $M$.
With the second-order moment of the distribution we have in principle 
a similar problem, although in this case the alternating 
series can be transformed into a non alternating one,
\be
<\bar{\zz}^2>=\frac{2}{M}
\sum_{l=1}^{M'} \left(\begin{array}{c}M'\\l\end{array}\right) 
\frac{(-1)^{l+1}}{l}=\frac{2}{M} \sum_{l=1}^{M'}\frac{1}{l}.
\label{eq:zz2}
\ee

For large values of $M'$ we can 
approximate the result using the definition of Euler's $\gamma$ constant,

\be
<\bar{\zz}^2> \approx \frac{2}{M} \left(\ln M'+\gamma \right).
\label{eq:eulerg}
\ee

Since the output of the procedure in absence of noise is

\be
\bar{\zz}_{\xx}=\sqrt{\rho_0} 
\frac{\sin (\pi \epsilon_0)}{N \sin (\pi \epsilon_0/N)},
\ee
the new {\it SNR\/} shall be (if $M'\gg 1$)

\be
\bar{\rho}=
\rho_0 \left(\frac{\sin (\pi \epsilon_0)}{N \sin (\pi \epsilon_0/N)}\right)^2
\frac{M}{2 \left(\ln M'+\gamma \right)}.
\ee

The standard procedure of averaging the square of the moduli of the DFTs 
gives for a general leaking signal the following {\it SNR\/}:

\be
\bar{\rho}_0=
\rho_0 \left(\frac{\sin (\pi \epsilon_0)}{N \sin(\pi \epsilon_0/N)}\right)^2
\frac{\sqrt{M}}{2},
\ee
what represents an improvement with the present method of

\be
\frac{\bar{\rho}}{\bar{\rho}_0}=\frac{\sqrt{M}}{\ln M' +\gamma},
\ee
a ratio larger than one when (\ref{eq:eulerg}) holds, which increases with
the number of blocks.

\section{Leaking signals embedded in unknown spectrum noise}

\subsection{A spectral estimator}

The task of replacing $S(k;M)$ in the filter definition is much more complex
than in the non-leaking case. The natural starting point is the value of 
\mbox{$\dtilde{\uu}(k;q/M')$} for $q$ other than $\bar{q}$, but this presents
two main problems. First of all, the fact that we are only able to choose
$\bar{q}$ out of a discrete set leaves open the unpleasant possibility that
the value of the signal frequency lies just in the middle of a bin. This
means that we shall ignore the precise way in which the signal energy will
be distributed among the different $q$'s, and thus the relative magnitude of
$\dtilde{\uu}(k;\bar{q}/M')$ when compared with the rest of outputs. The
scenario can be even worse, because we have no guarantees that $\bar{q}/M'$
corresponds to the best approach to the actual $\epsilon_0$: the noise can
mislead us into an inaccurate value of the signal frequency, leaving thus
the {\it true\/} $\dtilde{\uu}(k;\epsilon_0)$ among the discarded ones.

Instead of constructing an estimator for $S(k;M)$ in the hope that none
of the previously stated possibilities really takes place, what could
lead us again to a filtering procedure too sensitive to the signal's
peculiar properties, we are going to choose a {\it democratic\/} estimate
$\Bbb{S}(k;N)$: perhaps it will not be so accurate as it could, but
it will not show appreciable differences in its performance depending
on the actual frequency of the GW.

We define $\Bbb{S}(k;N)$ through an expression that closely resembles
that in equation (\ref{eq:BbbS}), 

\bea
\Bbb{S}(k;N) &\equiv& \frac{T}{N M} \frac{1}{M'-2} 
\sum_{q\neq\bar{q}} \left[
\left(\Re\{\dtilde{\uu}(k;q/M')\}
-\frac{1}{M'-1} \sum_{q'\neq\bar{q}} \Re\{\dtilde{\uu}(k;q'/M')\}\right)^2
\right. \nonumber \\
&+&\left.\left(\Im\{\dtilde{\uu}(k;q/M')\}
-\frac{1}{M'-1} \sum_{q'\neq\bar{q}} \Im\{\dtilde{\uu}(k;q'/M')\}\right)^2
\,\,\right],
\eea
where the $\tilde{\uu}_{\alpha}(k)$ have been substituted by
$\dtilde{\uu}(k;q/M')$, and the sums do not contain the term where
the signal is supposed to be located, $\dtilde{\uu}(k;\bar{q}/M')$.

Once more we replace $S(k;N)$ by $\Bbb{S}(k;N)$ in the definition of
$\bar{\zz}$, in order to get a new random variable $\Bbb{Z}$ which we
can compute using only experimental data. This quantity inherits two
characteristic traits from the way we estimate the spectral density
of the noise.
 
On the one hand, if the signal is large enough the filter output may have
a saturation limit, which will depend upon the particular values of some
parameters, such as $M\/$ or $M'$. This means that $\Bbb{Z}$ will not
surpass a certain threshold, even if the amplitude of the signal increases
indefinitely. The reason for that behaviour can be found in the fact that
when the signal is much more intense than the noise, what becomes really
difficult to assess is not the presence of the former but the properties
of the latter. So in these cases $\Bbb{S}(k;N)$ will overestimate $S(k;N)$.
It must however be stressed that the whole effect results in a change in
the value of $\rho_0$, and thus does not actually set an upper bound in
the {\it SNR\/} $\bar{\rho}$. On the other hand, when no signal is present
at a particular frequency, $\Bbb{S}(k;N)$ will underestimate $S(k;N)$,
since we do not use $\bar{q}$ when computing it, and
$\dtilde{\uu}(k;\bar{q}/M')$ has the biggest modulus among all the
$\dtilde{\uu}(k;q/M')$.

To compute the overall probability density of $\Bbb{Z}(k)$ is very difficult,
both in a single step (even if no signal is present) or in two steps, like
in Section \ref{se:lu}, since the auxiliary random variable $\Bbb{W}(k)$
defined like in (\ref{eq:W}) is no longer independent of $\bar{\zz}(k)$. We
shall thus obtain it by a different method. We introduce a {\it tentative\/}
$p(\Bbb{Z})$ inspired by the limiting probability density $p(\bar{\zz})$,
because when $M\/$ is large enough both must coincide:

\be
p(\Bbb{Z})=M M' w\Bbb{Z} e^{-\frac{M}{2}w \Bs{Z}^2} 
\left[1-e^{-\frac{M}{2} w\Bs{Z}^2}\right]^{M'-1}.
\ee

The new parameter $w\/$, which condenses all the differences between
$p(\bar{\zz})$ and $p(\Bbb{Z})$, measures in some suitable sense the
{\it bias\/} of the spectral estimator $\Bbb{S}(k;N)$, i.e.,

\be
w \sim  \frac{<\Bbb{S}(k;N)>}{S(k;N)}.
\ee

This point of view is somewhat vain since we are not in a position to
compute $<\Bbb{S}(k;N)>$ theoretically, as just stated. The value of $w\/$,
however, can be estimated from the filter output itself, using for instance
the relationship (\ref{eq:zz2}), replacing $\bar{\zz}$ by $\sqrt{w} \Bbb{Z}$,
and the statistical mean by an average over the filter output:

\be
w = \frac{\displaystyle 
N \sum_{l=1}^{M'} l^{-1}}{\displaystyle M \sum_{k=0}^{N/2-1} \Bbb{Z}^2(k)}.  
\label{eq:w}
\ee 

The procedure we shall use to compute (\ref{eq:w}) also reminds us that
$\Bbb{Z}(k)$ can be obtained from the experimental time series, irrespective
of its probability density. We set the value of the $w\/$ quantity {\it a
posteriori\/}, once the filtering process has finished. In fact, the
functional form of $p(\Bbb{Z})$ has its very origin in the comparative study
of the actual distribution $\Bbb{Z}(k)$ and the probability density of
reference, $p(\bar{\zz})$, for different values for the free parameters.
We pursue these ideas in the next section.

\subsection{A single filtering process}

\begin{figure*}
\begin{tabular}{lll}
\hspace*{-0.5 cm} & (a)&(b)\\
\hspace*{-0.5 cm} & \mbox{\psboxto(6.4cm;0cm){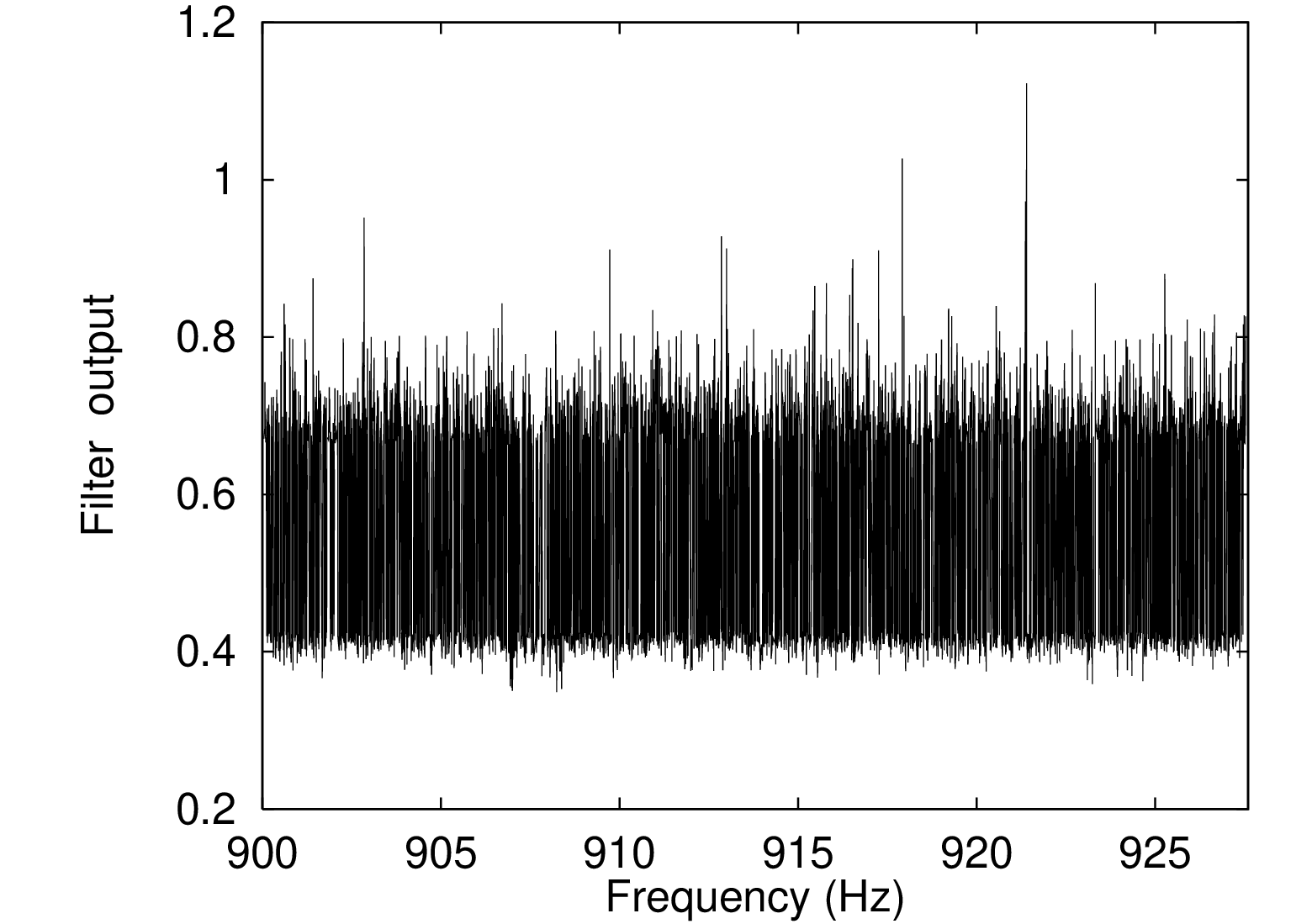}}&
\mbox{\psboxto(6.4cm;0cm){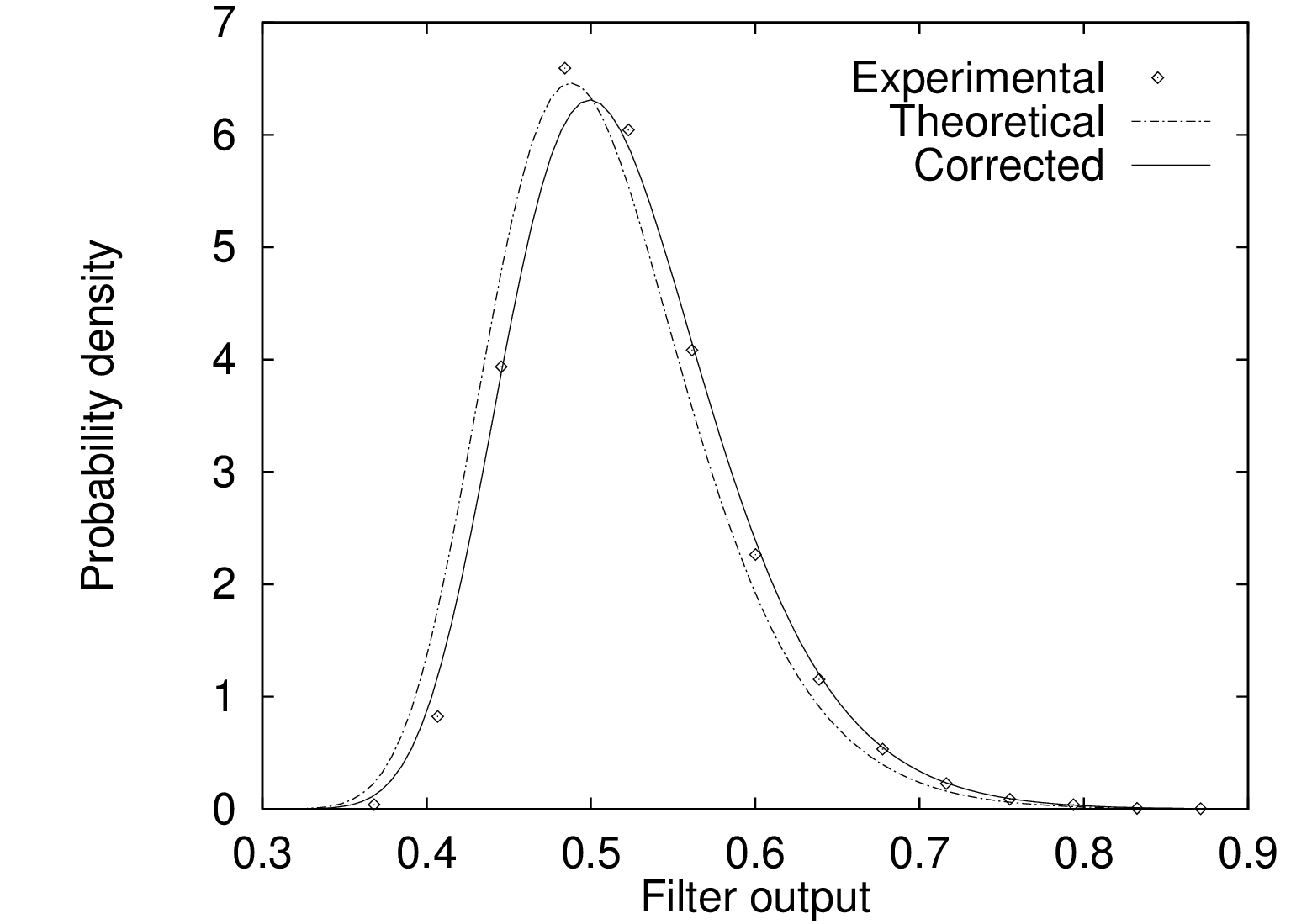}}
\end{tabular}
\caption{Here we find (a) the output of the filtering procedure, $\Bbb{Z}(k)$,
for $N=131072$, $M=36$ and $M'=64$, what represents about one day of 
data, and (b) its experimental distribution
compared with the theoretically expected, $p(\bar{\zz})$, and corrected 
one, $p(\Bbb{Z})$.   \label{fig:zz36}}
\end{figure*}

Let us begin analysing a stream of about one day of the {\it Explorer\/} data
with the layout used in Section \ref{se:exam}, i.e. $N\/$\,=\,131072 and
$M\/$=\,36\, $M'\/$\,=\,64. In fact, we will choose exactly the same time
series, starting on August 3rd of 1991, in order to be able to compare the
non-leaking and the leaking methods. So we get the results shown in Figure
\ref{fig:zz36}.a, where we have also externally introduced a signal, with
similar characteristics with respect to the previous example. Its amplitude
is now slightly bigger, $h_0$\,=\,3$\times$10$^{-23}$, i.e., {\it SNR\/} is
about 1, because the new method is slightly less sensitive than the
non-leaking one, as we have shown. And, of course, the signal spreads across
different frequency bins: $k_0=50918$ and $\epsilon_0$\,=\,0.1.

A plot of the distribution of $\Bbb{Z}(k)$ is displayed in Figure
\ref{fig:zz36}.b, where it is contrasted with $p(\bar{\zz})$ in
(\ref{eq:pzz}), the {\it theoretical\/} probability density, and with
$p(\Bbb{Z})$ once $w\/$ was computed following the prescription shown
in (\ref{eq:w}), the {\it corrected\/} one. The agreement of the latter
with the experimental probability density is again remarkable.

\begin{figure*}
\begin{tabular}{lll}
\hspace*{-0.5 cm} & (a)&(b)\\
\hspace*{-0.5 cm} & \mbox{\psboxto(6.4cm;0cm){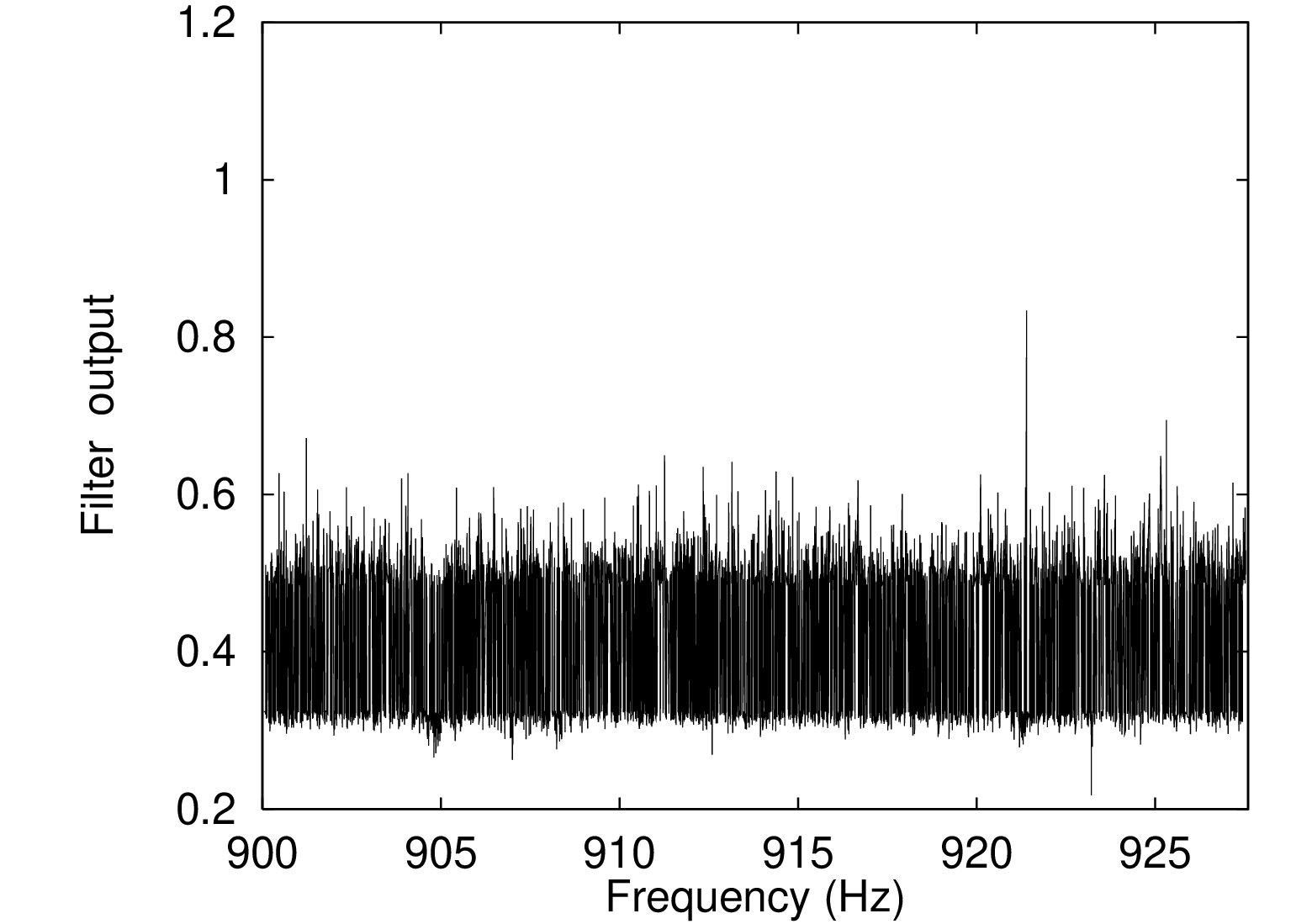}}&
\mbox{\psboxto(6.4cm;0cm){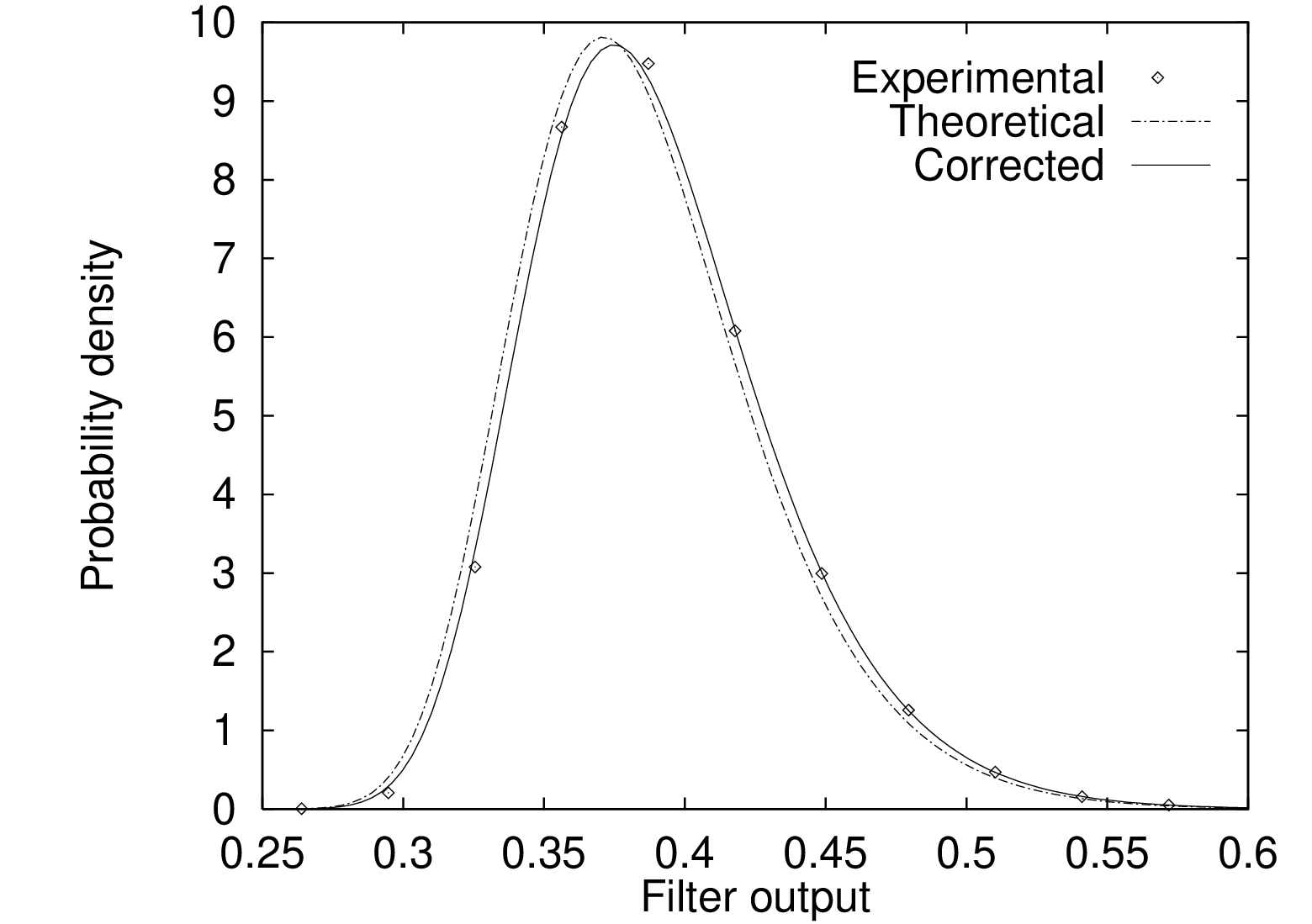}}
\end{tabular}
\caption{Here we find (a) the output of the filtering procedure, $\Bbb{Z}(k)$,
for $N=131072$, $M=72$ and $M'=128$, what represents about two days of
data, and (b) its experimental distribution compared with $p(\bar{\zz})$ and
$p(\Bbb{Z})$: in this case it is hard to distinguish from the one another.
\label{fig:zz72}}
\end{figure*}

It is convenient to point out that when the number of processed blocks 
increases, the value of $w\/$ rapidly approaches one, thus becoming an
irrelevant parameter. As a matter of fact, Figure \ref{fig:zz72}, where 
we present the same procedure with the same data, but extending the 
processed time to two days, shows that $p(\bar{\zz})$ is then a sufficiently
accurate expression for the probability density of the actual filter output. 

In the present case we can also compute the error of the first kind in terms
of the threshold $\lambda_0$:

\be
{\cal Q}_0 =
1-\left[1-e^{-\frac{M}{2} w\lambda_0^2}\right]^{M'},
\ee
and to invert this relationship,

\be
\lambda_0= \sqrt{-\frac{2}{M w} \ln 
\left(1-\left[1-{\cal Q}_0\right]^{\frac{1}{M'}}\right)}.
\ee

By way of example, the threshold levels in units of the graphs in Figures
\ref{fig:zz36} and \ref{fig:zz72} are 0.96 and 0.68, respectively, for a
false alarm probability ${\cal Q}_0$\,=\,10$^{-5}$. In either case the
signal is clearly above these thresholds, as it has respective heights of
1.12 and 0.83. From here we can rather accurately determine $\epsilon_0$,
too: we find $\bar q$\,=\,6 when $M'\/$\,=\,64 (i.e.,
$\epsilon_{0,{\rm estimated}}$\,=\,0.094), and $\bar q$\,=\,13 when
$M'\/$\,=\,128 (i.e., $\epsilon_{0,{\rm estimated}}$\,=\,0.101) for a real
value of $\epsilon_0$\,=\,0.1. 

\section{Outlook}

The long term operation of cryogenic GW detectors opens the possibility of
looking for long duration GW signals in the data generated by them, since
signal-to-noise ratios are enhanced by the availability of long integration
times. Monochromatic, as well as stochastic signals belong in this
category, though the latter require data from two or more independent
antennae. In this paper we have addressed the problem of the design of
suitable algorithms to single out possible monochromatic signals, coming
from any direction in the sky, in the background of the noisy data produced
by a cylindrical bar.

This is not such a simple problem, due to a variety of reasons: computers
are not arbitrarily powerful, detector duty cycles are not 100\%, data
quality is not homogeneous, Doppler shifts distort monochromaticity, etc.
Although some of the characteristics of the data are peculiar to the detector
system, there are a number of procedures which should be quite generally
usable. This paper is concerned with the problem to set up filtering
algorithms which enable the selection of candidate signals on the basis of
threshold crossing criteria of the filtered data. To this end we have
considered banks of filters which have a sinusoidal form but which enable
signal phase estimation. We have determined the pdf's at filter output and
thence consistently established the probability of crossing a given level.
We have also designed suitable estimators of the noise spectral density
which take into consideration the possibility that the signal be in one
or more of the FFT frequency bins for the complete duration of the
analysed data.

These procedures have been checked by means of simulations on top of real
data generated by the {\it Explorer\/} detector of the Roma group in 1991,
and they work quite well: the theoretical predictions on the improvement in
{\it SNR\/} as longer stretches of data are processed remarkably coincide with
those observed in the real data analysis. They thus appear promising, and
we would expect them to be useful to analyse other resonant detector data,
whether cylinders or the future projected large spherical antennae, and also
the large scale interferometers currently under construction, with suitable
adaptive modifications.

As we have seen with the simulations in this paper, the {\it Explorer\/}
detector comfortably sees amplitudes of 10$^{-23}$, and therefore one can
expect it to be sensitive to signals a few times 10$^{-24}$ ---see also
\cite{as97}. These are still rather high values for the expected amplitudes
of signals of astrophysical origin, so threshold crossing criteria must be
established having this in mind: if we set a very low false alarm tolerance
then chances of missing the real event will grow, whereas if we set it rather
low, the real signal will be treated on the same footing as random noise.

Real signals however have characteristic Doppler patterns, whose correct
assessment should provide a more sound selection of candidate signals. In a
forthcoming article we plan to apply the methodology here developped to a
systematic analysis of the above mentioned {\it Explorer\/} detector data
over a long period of time, and to extend it to also include Doppler shift
effects in the data. The very algorithms of section \ref{se:leak} will be
a powerful tool in this respect, if implemented under the perspective of
the zoom transform \cite{yip}, to enhance the frequency resolution within
the spectral neighbourhood of pre-selected candidate lines.

\section*{Acknowledgments}

The authors are grateful to the members of {\sl ROG\/} group at Rome for
giving them access to the data of the antenna {\it Explorer\/}, as well
as the detector's transfer function details and procedures related to
specific issues of the data analysis. We also thank Guido Pizzella for his
critical reading of the manuscript and valuable suggestions, and for his
continued hospitality. Bernard Schutz's comments, pointing us to the zoom
transform algorithm, have been very helpful and are greatly appreciated,
too. MM wishes to thank the DGR of the {\it Generalitat de Catalunya\/}
for financial support, and JAL the Spanish Ministry Education for a grant,
PB96-0384. We also acknowledge support from the {\it Institut d'Estudis
Catalans\/}.

\bsp

\label{lastpage}

\end{document}